\documentclass[nofootinbib]{revtex4-2}

\usepackage[english]{babel}

\usepackage[a4paper,top=2cm,bottom=2cm,left=3cm,right=3cm,marginparwidth=1.75cm]{geometry}

\usepackage{amsmath}
\usepackage{amssymb}
\usepackage{graphicx}
\usepackage[colorlinks=true, allcolors=blue]{hyperref}
\usepackage{color}
\usepackage{xcolor}
\usepackage{comment}
\usepackage{orcidlink}

\begin{document}

\title{Primary black-hole scalar charges and kinetic screening in $K$-essence-Gauss-Bonnet gravity}

\author{Guillermo Lara\,\orcidlink{0000-0001-9461-6292}}
	\email[]{glara@aei.mpg.de}
	\affiliation{Max Planck Institute for Gravitational Physics (Albert Einstein Institute), D-14467 Potsdam, Germany.}

    \author{Georg Trenkler\,\orcidlink{0009-0009-5206-6865}}
	\email[]{gtrenkler@ibs.re.kr}
	\affiliation{CEICO, Institute of Physics of the Czech Academy of Sciences, Na Slovance 1999/2, 182 00, Prague 8, Czechia}
    \affiliation{
    Cosmology, Gravity, and Astroparticle Physics Group, Center for Theoretical Physics of the Universe, Institute for Basic Science (IBS), Daejeon, 34126, Korea.}
 
	\author{Leonardo G. Trombetta\,\orcidlink{0000-0001-7345-5100}}
	\email[]{leonardo.trombetta@ib.edu.ar}
	\affiliation{CEICO, Institute of Physics of the Czech Academy of Sciences, Na Slovance 1999/2, 182 00, Prague 8, Czechia.}
    \affiliation{Centro At\'omico Bariloche and CONICET,\\ 
    S.C. de Bariloche, R\'io Negro, R8402AGP, Argentina}

\begin{abstract}
Black holes beyond General Relativity may carry non-standard charges that impact their phenomenology. We study how the scalar charge that is induced by the scalar-Gauss-Bonnet coupling is affected by the presence of a nontrivial kinetic term $K(X)$ in a test-field approximation. We discuss the corresponding kinetic screening in the asymptotically flat, static solution first. We then turn to the case where self-accelerating cosmology is driven by $K(X)$, finding that the time-dependence of the scalar field opens up the parameter space, turning the black-hole scalar charge from \emph{secondary} to \emph{primary}. We provide a stability analysis and a measure of the intensity of the kinetic screening from the quartic dispersion relation of the mixed scalar and gravitational modes.    
\end{abstract}

\maketitle

\section{Introduction}

The nature of the dark sector and the origin of cosmic acceleration remain open problems in cosmology, motivating the exploration of scalar-tensor modifications of gravity. This effort has led to a broad class of theories, including $K$-essence \cite{Armendariz-Picon:1999hyi, ArmendarizPicon:2000ah, ArmendarizPicon:2000dh, Garriga:1999vw}, Galileons \cite{Nicolis:2008in, Deffayet:2009wt}, and generalized Galileons \cite{Deffayet:2009mn, Deffayet:2010qz}, which feature derivative self-interactions and, in some cases, higher-order operators. It was subsequently realized that all these models fall within the Horndeski class \cite{Horndeski:1974wa, Kobayashi:2011nu}, the most general scalar-tensor theory with second order field equations, later extended to include degenerate higher-order scalar-tensor theories that avoid the propagation of additional degrees of freedom (d.o.f.) \cite{Zumalacarregui:2013pma, Gleyzes:2014dya, Langlois:2015cwa, Motohashi:2014opa} ---for reviews see Refs.~\cite{Langlois:2018dxi, Kobayashi:2019hrl}. On cosmological scales, such theories can account for late-time acceleration while remaining compatible with ---or even seemingly being favored (e.g. DESI \cite{DESI:2025zgx}) by--- observational constraints. In particular, the propagation of perturbations, even in anisotropic/nonlinear background configurations that might spontaneously break Lorentz invariance, is governed by an effective acoustic metric \cite{Sawicki:2024ryt}. The consistency and viability of these models are therefore tightly constrained by stability requirements, including the absence of ghost and gradient instabilities which can be formulated as conditions on the invariants of the acoustic metric, which play a central role in delineating the allowed parameter space, e.g. Ref. \cite{Creminelli:2019kjy}.

In the strong-field regime, static and stationary black holes (BHs) in General Relativity (GR) are remarkably simple, being uniquely described by the Schwarzschild and Kerr solution respectively in vacuum~\cite{Robinson:1975bv,Israel:1967wq,Hawking:1971vc}. With the exception of electromagnetism, such black holes carry no additional charges for Standard Model fields --- a result formalized in a list of classical no-hair theorems with varying assumptions
\cite{Hawking:1972qk,Bekenstein:1995un,Sotiriou:2011dz,Graham:2014mda, Graham:2014ina}, especially in the case of shift-symmetric theories \cite{Hui:2012qt}. The latter are favored as they are generically more robust against quantum corrections \cite{Pirtskhalava:2015nla,Santoni:2018rrx}. In most scalar-tensor theories, these no-hair results continue to hold for static configurations with asymptotically trivial boundary conditions. 
However, exceptions can arise, for example: \emph{i)} when the scalar couples to the Gauss-Bonnet term (GB) \cite{Sotiriou:2013qea,Sotiriou:2014pfa,Creminelli:2020lxn}, as generically anticipated within an effective field theory (EFT) framework, or \emph{ii)} when the scalar field deviates from a static configuration and/or asymptotically flat boundary conditions at infinity do not apply, among other types of exceptions. The first solution with a time-dependent scalar field while retaining a static metric was found in Ref.~\cite{Babichev:2013cya}. In general, however, such time-dependence introduces accretion or a nontrivial cosmological evolution, e.g. \cite{Horbatsch:2011ye,Babichev:2025ric}. In these cases, BHs can acquire scalar “hair”, either through direct sourcing by spacetime curvature or via a cosmologically induced timelike gradient at infinity where the resulting configurations connect the local BH environment to the global cosmological dynamics. These situations can modify intrinsic BH properties such as scalar charges, horizon structure, the innermost stable circular orbit (ISCO), and the Quasi-Normal-Mode (QNM) spectrum.

Screening mechanisms, including Vainshtein \cite{Vainshtein:1972sx}, 
kinetic screening/k-mouflage \cite{Babichev:2009ee} or Chameleon \cite{Khoury:2003aq} effects, play a central role in rendering scalar-tensor theories phenomenologically viable. By suppressing scalar-mediated forces at short scales, they reconcile modified gravity phenomenology with solar-system and laboratory tests of gravity \cite{Will:2014kxa}. The interplay between screening and cosmologically induced scalar gradients is particularly relevant for shift-symmetric theories, where derivative interactions can dominate near compact objects while cosmological evolution drives the background scalar field at large distances, both in deeply-nonlinear regimes. While this duality enables the existence of smooth, non-singular solutions interpolating between BH and cosmological horizons, though challenges remain in constructing solutions that transition between different branches consistently \cite{Babichev:2012re, Babichev:2025ric}. Furthermore, the associated non-linearities make the study of two-body systems significantly more complicated \cite{deRham:2012fw,Dar:2018dra}.

One screening mechanism that is particularly relevant for this work is kinetic screening, which has been extensively studied in the context of matter sources, like stars, neutron stars (NS), and solar-system-like objects. In such setups, recent works have emphasized that obtaining reliable predictions for the two-body problem within the post-Newtonian framework becomes highly nontrivial due to screening, especially in the nonlinear regime and for comparable-mass binaries \cite{Cayuso:2024ppe, Kuntz:2019plo, Boskovic:2023dqk}, e.g. binary pulsars greatly constrain Damour-Esposito-Far\` ese models~\cite{Kramer:2021jcw}, but such constraints critically rely on precise PN predictions.
The situation is relatively simpler for mixed black hole-neutron star (BHNS) binaries, where analytic progress can be made by exploiting the isolated solution for the screened compact object \cite{Boskovic:2023dqk, Cayuso:2024ppe}. Earlier attempts to model screening effects in the two-body problem, reviewed for instance in the conclusion of Ref.~\cite{Boskovic:2023dqk}, often rely on additional assumptions or constructions, such as the introduction of fictitious bodies, which introduce conceptual and practical caveats \cite{deRham:2012fw, Dar:2018dra, Hertzberg:2022bsb}.

Recent theoretical work has highlighted that these cosmologically induced scalar configurations can lead to observable consequences for gravitational wave signals from BH binaries and may constrain the coefficients of higher-order interactions in EFT frameworks. The inclusion of multiple energy scales in the scalar-tensor action even further enriches the phenomenology \cite{Noller:2019chl,Thaalba:2025lwe}. Overall, the study of scalar-tensor black holes embedded in cosmological backgrounds provides a bridge between the physics of the dark sector and strong-field gravitational phenomena, offering a unified arena to explore the implications of modified gravity from cosmology down to horizon-scale astrophysics.

In this work we follow this idea and aim to provide a concrete realization within a simple enough model, which however is sufficiently rich to offer nontrivial phenomenology in cosmology and around BHs. For this purpose we will consider a $K$-essence type scalar-tensor theory, with the addition of the linear scalar-Gauss-Bonnet operator
\begin{equation} \label{action}
    S = M_P^2 \int d^4 \! x \sqrt{-g} \left[ \frac{R}{2} + K(X) + \alpha \phi \, \mathcal{G} \right] + S_m[\psi, e^{4 \beta \phi}g_{\mu\nu}] \, ,
\end{equation}
where $X = - \frac{1}{2} g^{\mu\nu} \partial_\mu \phi \partial_\nu \phi$ is the kinetic term of the scalar field, \mbox{$\mathcal{G} = R_{\mu\nu\rho\sigma} R^{\mu\nu\rho\sigma} - 4 R_{\mu\nu} R^{\mu\nu} + R^2$} is the Gauss-Bonnet invariant, $S_m$ is the matter action, and \(\psi\) are the matter fields. The coupling constant $\alpha$ has dimensions of length squared. We have included a conformal coupling between matter and the scalar field, with coupling strength $\beta$, to make a connection to the usual study of kinetic screening around matter sources given in the literature, however our focus will be on black holes.

In the absence of a scalar-matter coupling $\beta$
the theory enjoys a shift symmetry $\phi \to \phi + c$, and the scalar equation of motion (EoM) takes the form of a covariant conservation of the associated current $J^\mu$,
\begin{eqnarray} 
    \nabla_\mu J^\mu = 0 \, .
\end{eqnarray}
For the theory at hand, this explicitly reads
\begin{eqnarray} \label{eq: scalar equation covariant}
\nabla_{\mu}[K_{X} \nabla^{\mu} \phi] = - \alpha  \mathcal{G} \, .
\end{eqnarray}
The corresponding equations for the metric read
\begin{eqnarray}\label{eom-metric-bg}
    G_{\mu\nu} &= K_X \partial_\mu\phi \partial_\nu\phi + K g_{\mu\nu} - \alpha (g_{\rho\mu}g_{\delta\nu} + g_{\rho\nu}g_{\delta\mu}) \nabla_\sigma\left(\partial_\gamma \phi \: \epsilon^{\gamma\delta\alpha\beta}\epsilon^{\rho\sigma\lambda\eta} R_{\lambda\eta\alpha\beta}   \right) + M_{P}^{-2} T^{(m)}_{\mu \nu}\, ,
\end{eqnarray}
where $G_{\mu\nu}$ is the Einstein tensor, $R_{\lambda\eta\alpha\beta}$ the Riemann tensor, and \(T^{(m)}_{\mu\nu} \equiv -(2/\sqrt{-g})\delta S_m/\delta g^{\mu\nu}\) the matter energy-momentum tensor in the Einstein frame.

The presence of the $\alpha$ term acts as a source for the scalar field whenever the curvature of the spacetime is nontrivial, even when there is no coupling to matter. While this is both the case in cosmology and around black holes, the length scales at which its effect is relevant are determined by $\sqrt{\alpha}$. We will assume in what follows that $\sqrt{\alpha} \ll H^{-1}$, making the effect of this term negligible in cosmology, in which case the standard $K$-essence solutions are valid up to perturbative corrections $\mathcal{O}(H^2\alpha)$. Instead, at shorter astrophysical scales, which we identify generically with the Schwarzschild radius $r_s$ of a black hole, this source term will be responsible for generating a nontrivial scalar profile with an associated dimensionless scalar charge $\alpha_\text{BH} \sim \alpha/r_s^2$. This is a \emph{secondary} charge, as it is uniquely fixed to the mass of the black hole and not a free parameter of the solutions. Current experimental bounds based on the absence of an observed dephasing of the gravitational waveform due to scalar-wave emission, put a constraint on the charge of order $\alpha_\text{BH} \lesssim 10^{-2}$ for black-holes observed by LVK \cite{Lyu:2022gdr, Sanger:2024axs}. This bound can only be translated to $\alpha$ directly when no significant screening effect is present, otherwise the actual bound on $\alpha$ may be much looser \cite{Noller:2019chl,Thaalba:2025lwe}, as only the effective charge is being constrained. This theory has a non-canonical kinetic term $K(X)$ with the added requirement of sustaining self-accelerating solutions, and therefore it is inherently nonlinear. This is expected to give rise to a strong screening of the scalar charge at short distances.  

The outline of the paper is as follows: In Sec. \ref{sec:flat}, we review kinetic screening for NS and extend the analysis to BHs in asymptotically-flat spacetimes, where we study near horizon behaviour. In Sec. \ref{sec:CosmoBHs}, we discuss the cosmological background solutions as well as BH background solutions and investigate the physics at the horizons in order to determine the regularity conditions that fix the value of the scalar charge $\alpha_\text{BH}$. These conditions seem to be flexible enough to allow for $\alpha_\text{BH}$ to be a free parameter ---a \emph{primary} charge. In Sec. \ref{sec:eikonal}, we study perturbations on top of these backgrounds. For this we employ the eikonal approximation to de-mix the system of characteristic equations as much as possible, and then define a unique, characteristic invariant encoding information about stability and screening. We study this quantity in two simple limits to demonstrate that it factorizes into products of acoustic metrics there. In Sec. \ref{sec:example} we show our analytic test-field solutions, explain which branches are relevant and use the characteristic invariant to study both stability and screening of these solutions. In this way we constrain the corresponding parameter space, finding that the scalar charge still remains of \emph{primary} type. Finally, in Sec. \ref{sec:conclusion}, we summarize our results and provide an outlook for future work.

\section{Kinetic screening in asymptotically-flat spacetimes}\label{sec:flat}

\subsection{Matter sources}

We first consider the case where \(\alpha = 0\) and the scalar field is conformally-coupled to matter in the action. This is the scenario in which kinetic screening is often discussed in the literature ---see e.g.~Refs.~\cite{Babichev:2009ee, Brax:2012jr, Brax:2014gra, terHaar:2020xxb,Bezares:2021yek, Bezares:2021dma, Lara:2022gof,Boskovic:2023dqk, Cayuso:2024ppe}.
The scalar equation of motion is then of the form
\begin{equation}
    \nabla_{\mu}[K_{X} \nabla^{\mu} \phi] =  \dfrac{\beta}{M^{2}_P} \, T ~,
\end{equation}
where \(T\) is the trace of the matter stress-energy tensor. 

We now illustrate the main features of kinetic screening in a simple set up. Following the discussion in \cite{Bezares:2020wkn}, we consider a pointlike matter source \(T = - m \, \delta^{(3)}({\bf r})\)
in flat space, where \(\delta^{(3)}({\bf r})\) is a delta distribution, and \(m\) is the particle mass in the Einstein frame. The latter differs from the mass in the Jordan frame, where matter is minimally coupled, as $m_J = e^{-2\beta \phi} m$. We also choose \(K(X) = X + c_2 X^2/M^2\), where \(c_2 \sim O(1)\) is a dimensionless constant and \(M\) has units of inverse length. 
Integrating the scalar equation once yields
\begin{align}\label{eq: scalar equation matter}
    \phi' - \dfrac{c_2}{M^2} {(\phi')}^3 = -\dfrac{\beta m}{4 \pi M^{2}_{P} r^2}~.
\end{align}
At large separations, \(\phi' \ll  M \), and the cubic term can be neglected. Then, the solution in the linear regime is well-approximated by
\(\phi' \simeq (-\beta m / 4 \pi M^{2}_{P}) r^{-2}\).
Close to the source, instead, \(\phi' \gg M \) and the cubic term dominates. The solution is well-approximated in this regime by the solution to a nonlinear cubic equation, 
\(\phi' \simeq [(M^2/c_{2}) (\beta m / 4 \pi M^{2}_{P})]^{1/3} r^{-2/3}\).
Having a continuous solution connecting both regimes requires that \(c_2 < 0\). Notice, however, that positivity bounds from scattering amplitudes require instead \(c_2 > 0\) in order for this flat-space model to admit a standard UV completion~\cite{Adams:2006sv}.
The boundary between such linear/non-linear regimes, or the screening radius \(r_k\), can be estimated by demanding in Eq.~\eqref{eq: scalar equation matter} that both the linear and cubic terms are of the same order, i.e. when \(\phi' \sim M \).
We find \(r_{k} \sim ( m / M^{2}_{P} M )^{1/2}\).
In the screening region \(r \ll r_k\), the scalar gradient (the ``fifth force''), \(\phi' \sim (r/r_k)^{4/3} \phi'_\text{lin}\), is highly suppressed with respect to the would-be linear behaviour \(\phi'_\text{lin} \sim r^{-2}\) if there were no nonlinear operators in \(K(X)\).
Outside the screening region, we can associate a scalar charge
\(\alpha_\text{NS}= \beta /(8\pi M^2_{P})\), 
defined by the asymptotic behaviour of the scalar field, 
\(\phi(r \to \infty) = \phi_\infty + 2 m \alpha_\text{NS}/r+O(r^{-2})\)~.
From Eq.~\eqref{eq: scalar equation matter},  we also see that the scalar field is trivial (\(\phi = \text{const.}\)) in the vicinity of BHs, a fact encapsulated in several no-hair theorems~\cite{Hui:2012qt, Graham:2014mda, Graham:2014ina}.
Therefore no screening for BHs can occur in this situation when \(\alpha=0\).

\begin{figure}[h]
    \centering
    \includegraphics[width=2.8in]{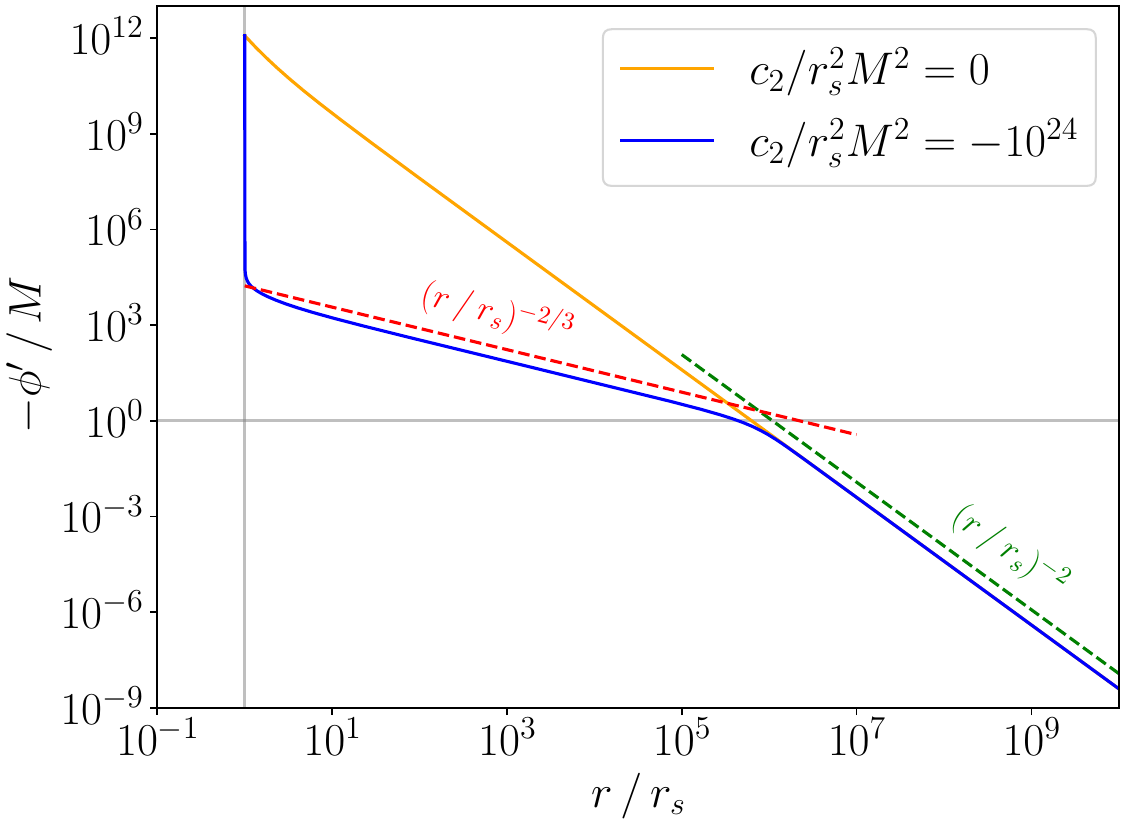}
    \caption{\emph{Screening for asymptotically-flat black holes.}
    We plot (blue) the normalized scalar gradient for a solution featuring screening in a Schwarzschild BH background. We set \(\alpha_\text{BH} = 4\alpha/r_s^2= 2/5 \) and \(c_2/r_s^2 M^2 = -10^{24}\) for illustrative purposes.
    The screening radius is located at \(r_k/r_s \simeq 3 \times 10^5\) and corresponds to the knee in the blue line.
    In dashed lines, we show guiding slopes with the expected behaviour.
    For reference, we also plot (orange) a solution where there are no nonlinear operators to make the gradient suppression apparent.
    At the location of the BH horizon (vertical line), both solutions coincide.
    }
    \label{fig: BH screening asymp flat}
\end{figure}

\subsection{Black holes} \label{sec: screening for bhs asympt flat}

A linear coupling to the Gauss-Bonnet invariant (\(\alpha \neq 0\)) evades no-hair theorems and leads to a non-vanishing scalar source term in vacuum~\cite{Sotiriou:2013qea, Sotiriou:2014pfa}.
To illustrate this, consider a Schwarzschild BH of mass \(m\) and Schwarzschild radius \(r_s \equiv 2m\).
Continuing with a point-particle description, we can approximate \(\mathcal{G} \simeq 16 \pi r_{s}^{-1} \delta^{(3)} (\mathbf{r})\). (When integrated over \(\mathbb{R}^3
\), this approximation yields the same value as integrating the Gauss-Bonnet invariant corresponding to Schwarzschild, \(\mathcal{G} = 48 m^2/r^6\), in the region outside the BH.)
Integrating once, the scalar equation~\eqref{eq: scalar equation covariant} becomes
\begin{align} \label{eq: scalar equation vacuum point particle}
    \phi' - \dfrac{c_2}{M^2} {(\phi')}^3 = -\dfrac{4\alpha }{r_{s} r^2}~.
\end{align}
In complete analogy with the matter case, we see that screening for BHs can occur in the region \(r \ll r_k\), where the screening radius is now given by 
\(r_k \sim [\alpha / (r_s M)]^{1/2}\)~.
In the interior region, \(r \ll r_k\),
the scalar field behaves nonlinearly,
\(\phi' \simeq [(M^2 /c_2)(4 \alpha/r_s)]^{1/3} r^{-2/3}\).
Whereas in the exterior region, \(r \gg r_k\),
we recover the well-known linear behavior of the scalar field,
\(\phi' \simeq (-4\alpha/r_s) r^{-2}\), which is reminiscent of the theory without non-linear terms in \(X\). Notice that we require again \(c_2 < 0\).
Finally, in the exterior region, we can extract a unique value for the scalar charge \(\alpha_\text{BH}=4\alpha/r_s^2\),
defined by the asymptotic behaviour of the scalar field, \(\phi(r \to \infty) = \phi_\infty + \alpha_\text{BH}r_s/r+O(r^{-2})\). This is a \emph{secondary} charge, as it is fixed by the coupling $\alpha$ and the mass of the black hole $m$.

\subsection{Behaviour near the origin/horizon}

While the point-particle description is convenient to describe the qualitative behaviour of screened solutions, this approximation is not able to capture the behaviour of the scalar field near the origin for a matter source or near the horizon for a BH.
A slightly more realistic description is to represent the energy density \(\rho\) of a star as an extended distribution, e.g. \(T \simeq - \rho(r)\), with \(\rho(r) \equiv \max\{\rho_{0} [1-(r/r_\star)^2],  0\}\), \(\rho_0 = \text{const.}\) and star radius \(r_\star\).
Since \(\phi' \to 0\) at the center of the star, the linear behaviour becomes dominant as \(r \to 0\).
Therefore, screening is not active in a small region near the center of the matter distribution ---see e.g.~Refs.~\cite{Brax:2012jr, terHaar:2020xxb, Lara:2022gof}.

For the BH case, we can improve our description by solving the scalar equation in the test-field approximation on a Schwarzschild background.
The spacetime is described by \(ds^2 = -f(r) \, dt^2 + [f(r)]^{-1} \, dr^2 + r^2 \, [d \theta^2 + \sin^2 (\theta) \, d\varphi^2] \), where \(f(r) \equiv 1 - r_s/r \)~.
Integrating the scalar equation once, we obtain
\begin{align} \label{eq: scalar equation vacuum dec lim}
    \left(1-\dfrac{r_s}{r}\right) \phi' - \dfrac{c_2}{M^2} \left(1 - \dfrac{r_s}{r}\right)^2\left(\phi'\right)^3 
    = -\dfrac{4\alpha}{r_s r^2} \left(1-\dfrac{r_s^3}{r^3}\right)~.
\end{align}
The scalar gradient is regular at the horizon.
We determine its value by expanding the scalar gradient around \(r_s\) as \(\phi'(r)=\phi_1 +\phi_2 (r-r_s) + \cdots\), and by solving Eq.~\eqref{eq: scalar equation vacuum dec lim} perturbatively in \(x \equiv r-r_s\). %
The first two coefficients are \(\phi'(r_s) = \phi_1 = -12 \alpha / r_s^3 \) and
\(\phi_2 =18 \alpha (r_s^6-48 \alpha^2 c_2  M^{-2})/r_s^{10}\)~.
Since \(\phi'(r_s)\) is independent of \(c_2\), it matches the solution we would have obtained had we considered only the linear term.
Similar to the matter case, we conclude that in this case screening is not active in the near vicinity of the horizon.
In general, the cubic equation~\eqref{eq: scalar equation vacuum dec lim} can be easily solved analytically with the aid of computer algebra software.
We plot an example solution in Fig.~\ref{fig: BH screening asymp flat}.
It seems plausible to extend these solutions beyond the decoupling limit. Indeed, Ref.~\cite{Thaalba:2022bnt} obtained solutions with backreaction in the presence of the \(c_2\)-term without looking for screening.

\section{Black holes in a self-accelerating universe}\label{sec:CosmoBHs}

\subsection{Homogeneous cosmology}

Let us look for self-accelerating solutions in cosmology in this model in a very simplistic approach that will be sufficient for our purposes ---other realizations are possible, see e.g.~Ref.~\cite{Brax:2014wla}. For simplicity we consider de Sitter space
\begin{eqnarray}
    ds^2 = - d\tau^2 + e^{2H\tau} \left( d\rho^2 + \rho^2 d\Omega^2 \right) \,.
\end{eqnarray}
In this case we assume the scalar field to be $\phi = \phi(\tau)$. Then the scalar equation has an attractor solution ($\ddot{\phi} = 0$) of the form
\begin{eqnarray} \label{homog-sol}
    \phi_c = q_c \, \tau \, ,    
\end{eqnarray}
with $q_c$ fixed by the condition of vanishing shift-charge density $J_0 = 0$,
\begin{eqnarray} \label{cosmo-eq}
    q_c K_X(X_c) = 8 \alpha H^3 \, ,
\end{eqnarray}
where $X_c = q_c^2/2$. The corresponding Friedmann equation reads
\begin{eqnarray} \label{Friedmann-eq}
    3 H^2 = 2X K_X - K - 24 \alpha q_c H^3 \,.
\end{eqnarray}
These equations have several branches of solutions. On the one hand, there is one where the sGB term acts as the main source for the scalar field, $q_c \simeq 8 \alpha H^3$ (assuming $|K_X| \sim 1$), but by our assumption $\sqrt{\alpha} \ll H^{-1}$ it is not good for driving self-acceleration. Moreover, in the $\alpha \to 0$ limit in fact this branch corresponds to a typically unstable Minkowski vacuum (assuming no cosmological constant) with no condensate $\phi_c = 0$ and ghost-like perturbations $K_X(0) < 0$.
On the other hand, the branches satisfying the approximate equation
\begin{eqnarray}
    K_X(X_c) \simeq 0 \,,
\end{eqnarray}
exhibit a condensate with stable perturbations. This mechanism is known as ghost condensation \cite{Arkani-Hamed:2003pdi}.

The cosmology therefore is approximately given by the usual $K$-essence type solutions. These are known to have a vanishing soundspeed $c_s$ of scalar perturbations \cite{Scherrer:2004au}, potentially suffering from strong-coupling problems. Here, provided the proper sign is chosen for $q_c$, i.e. $\alpha\, q_c > 0$, the soundspeed is nonzero and real $c_s^2 > 0$. However, again due to the assumed separation of scales, it is still very small and likely to still give rise to strong coupling. For this reason we shall consider these as toy models, but nevertheless still work with them due to their relative simplicity. More realistic models are left for future work. 

\subsection{Black holes embedded in cosmology}

We are now interested in studying the solution for the scalar field around a static, spherically symmetric metric describing a black-hole embedded in cosmology. In static coordinates we assume for the metric to be of the form
\begin{eqnarray} \label{metric-ansatz}
    ds^2 = - h(r) dt^2 + \frac{dr^2}{f(r)} + r^2 d\Omega^2\,,
\end{eqnarray} 
where in principle $f(r)$ and $h(r)$ may be different. In latter sections of this paper we will consider test-field solutions on top of a Schwarzschild-de Sitter background,
\begin{eqnarray}\label{SSdSfandh}
    f(r) = h(r) = 1 - \frac{r_s}{r} - H^2 r^2 \, ,
\end{eqnarray}
where $r_s$ and $H$ are the Schwarzschild radius and Hubble constant, respectively. This metric has two horizons when $f=0$ at positions slightly shifted from $r_s$ and $H^{-1}$, but parametrically close under the assumption of a large separation of scales $H r_s \ll 1$. In what follows we will keep the expressions general for arbitrary $f(r)$ and $h(r)$ when possible.

The correct solution for the scalar field in the presence of a compact source, such as a black hole, should approach the homogeneous one described above in the appropriate limit. It is useful then to express \eqref{homog-sol} in static coordinates of the de Sitter metric, for which $f = 1 - H^2 r_s^2$,
\begin{eqnarray} \label{cosmo-phi-static-coords}
    \phi_c = q_c \, t + \frac{q_c}{2H} \log \left[ 1 - (Hr)^2 \right]\, .
\end{eqnarray}
This form prompts that we consider an ansatz for the scalar field which has both radial and time dependence. For simplicity we may assume the latter to also be linear in the "local" time
\begin{equation} \label{scalar-ansatz}
    \phi = q\,t + \varphi(r) \, ,
\end{equation}
where we intentionally let $q$ be different from $q_c$. Due to the shift-symmetry, such linear time-dependence does not change the nature of the metric equations, which remain time-independent\footnote{However, the constant $q$ does appear in the equations nontrivially.}, making it a priori compatible with a static metric like \eqref{metric-ansatz}. There is, however, another source of possible time-dependence of the metric, i.e. accretion, which we will neglect on the assumption it can be made sufficiently small for the current model. We will check this assumption later.

The shift-symmetry once again allows the scalar field equation to be easily integrated once. Using the above ans\"atze it gives for the theory \eqref{action},
\begin{eqnarray} \label{scalar-eom}
    \frac{J_r}{M_P^2} = - K_X \, \varphi' + \frac{4\alpha h'}{r^2 h}(1-f) =  \frac{\alpha_\text{BH} r_s}{r^2} \frac{1}{\sqrt{h f}} \,,
\end{eqnarray}
where $\alpha_\text{BH}$ is an integration constant --- the scalar charge --- to be determined by regularity conditions at the horizon(s). A regularity condition may be posed, for example, on invariant quantities constructed from the derivatives of the scalar field. The simplest one is the kinetic term 
\begin{eqnarray}
    X = \frac{1}{2} \left( \frac{q^2}{h} - f \varphi'^2 \right)\,,
\end{eqnarray}
which is finite at the horizon $r = r_h$ provided
\begin{eqnarray} \label{near-horizon-varphip}
    \varphi' &=& - \frac{q}{\sqrt{h f}} + \psi'(r)\,,
\end{eqnarray}
where $\psi'(r_h)$ is finite and related to $X(r_h)$. This form also ensures other invariant quantities like $\square \phi$ to be finite, while the sign choice agrees with \eqref{cosmo-phi-static-coords} at the cosmological horizon, ensuring a finite $\phi$. A near horizon expansion using \eqref{near-horizon-varphip} and the scalar field equation \eqref{scalar-eom} reveals a regularity condition per each, the black-hole and cosmological horizons. Respectively, 
\begin{eqnarray}
    \alpha_\text{BH} &=&  K_X(X_\text{BH}) \, q r_s + \frac{4 \alpha}{r_s^2} \, , \label{reg-cond-BH} \\
    \alpha_\text{BH} &=&  K_X(X_\text{ch}) \, \frac{q H^{-1}}{(H r_s)} - \frac{8\alpha}{r_s^2} (H r_s)  \, . \label{reg-cond-cosmo}
\end{eqnarray}
Notice that in the absence of $q$, the scalar-field time derivative, it is not possible to satisfy both these equations simultaneously due to the large separation of scales $H r_s \ll 1$. This is the origin of the non-construction of fully static solutions as discussed in Ref.~\cite{Babichev:2024txe}. As also pointed out therein, a canonical choice $K(X) = X$ is also problematic as satisfying both the above conditions at the same time requires $|q| \gg |q_c|$, leading to a parametrically slow approach to homogeneity at large distances
\begin{equation} \label{approach-to-homogeneity}
    X \simeq X_c + e^{-2H\tau} \frac{(q-q_c)^2}{2(H \rho)^2} \,,
\end{equation}
where we have re-expressed \eqref{scalar-ansatz} in terms of Friedmann coordinates and taken the superhorizon limit $H \rho \gg 1$. In this limit equation \eqref{scalar-eom} approaches the cosmological one $K_X \simeq 0$, which universally gives $\varphi' \simeq q_c/Hr$.

A nontrivial $K(X)$ allows instead to solve Eqs.~\eqref{reg-cond-BH} and \eqref{reg-cond-cosmo} in any number of ways by including new variables into the mix, namely $X_\text{BH}$ and $X_{ch}$. A priori one may choose $q \simeq q_c$ and $\alpha_\text{BH} \sim \alpha/r_s^2$, but also any other combination, provided the equation of motion allows to connect the values at both horizons. This is a novel aspect of considering nontrivial kinetic operators in conjunction with the scalar-Gauss-Bonnet coupling and a time dependent scalar field: the scalar charge $\alpha_\text{BH}$ is no longer uniquely fixed in terms of the black-hole mass, $r_s$, but is in principle a free parameter. In other words, the scalar charge goes from \emph{secondary} to \emph{primary}. 

Other criteria may restrict the possible choices of $\alpha_\text{BH}$ and $q$ beyond the two regularity conditions at the horizons. This includes the actual existence of real solutions in the whole region in between that have the correct asymptotics, are sufficiently long lived, and are stable under perturbations. As alluded to above, accretion makes the solution never truly stationary, and is a consequence of diffeomorphism invariance that every time there is simultaneously non-vanishing time-derivative $q$ and scalar charge $\alpha_\text{BH}$, there is an energy flux into the black hole \cite{Babichev:2015rva},
\begin{equation}
\label{energy-flux}
T^r{}_t = - q \, J^r \,.
\end{equation}
This puts a bound on how large the combination $q \, \alpha_\text{BH}$ should be in order to trust the solution to be quasi-stationary on the astrophysical scales of interest. 

The question of stability requires more work, especially in light of the kinetic mixing of scalar and gravitational modes induced by the scalar-Gauss-Bonnet operator. As it is not possible to fully decouple these modes in general, a specialized approach to this question is prompted. This is the focus of the next section, where we will propose a necessary condition for stability, which while  not sufficient, it will still allow us to restrict the parameter space in an explicit example later in Sec~\ref{sec:example}. As an added benefit, our approach will also enable us to discuss screening, a fundamental aspect of any phenomenological analysis of the solutions.

\section{Perturbations and the eikonal limit}\label{sec:eikonal}
\subsection{Demixing the system of equations, dispersion relations and the characteristic invariant}

We study linear perturbations of the background equations of motion \eqref{eq: scalar equation covariant} and \eqref{eom-metric-bg} by decomposing the metric and scalar field as
\begin{equation}
g_{\mu\nu}=\bar{g}_{\mu\nu}+h_{\mu\nu} \qquad,\qquad
\phi=\bar{\phi}+\pi \, .
\end{equation}
The background metric $\bar{g}_{\mu\nu}(r)$ is taken to be of the form \eqref{metric-ansatz}, while the background scalar field $\bar{\phi}(t,r)$ is given by \eqref{scalar-ansatz}. Here $h_{\mu\nu}$ denotes the metric perturbation and $\pi$ the scalar fluctuation. For notational convenience, we will suppress the bars over background quantities in what follows.
Upon linearization, both the scalar and metric equations of motion contain second derivatives of $h_{\mu\nu}$ and $\pi$. To extract their leading-order (characteristic) structure, we employ the eikonal, or geometric-optics, approximation. Following e.g. \cite{LandavshitzII,Courant} (see also e.g. \cite{Barcelo:2005fc,Sawicki:2024ryt} in the context of acoustic geometry or \cite{Glampedakis:2019dqh,Silva:2019scu,Bryant:2021xdh} in the context of quasinormal modes beyond GR), we introduce a small parameter $\epsilon$ that controls the ratio between the wavelength of the perturbations and the characteristic length scale of the background and write
\begin{equation}
h_{\mu\nu}=\mathcal{A}^h_{\mu\nu} e^{i S/\epsilon} \qquad,\qquad
\pi=\mathcal{A}^\phi e^{i S/\epsilon}\; .
\end{equation}
This approximation is valid for rapidly oscillating perturbations whose wavelengths are much shorter than the scale over which the background varies. We further assume that the scalar and metric perturbations share a common phase $S$, and identify the associated momentum co-vector
\begin{equation}
p_\mu \equiv \partial_\mu S \, .
\label{eq:momentumCovector}
\end{equation}

The (gauge-reduced) characteristic polynomial $\mathcal{C}$ associated with the principal symbol of the total system decomposes into a purely gravitational (axial) part and a mixed (polar) scalar-gravitational part: $\mathcal{C}(p)=Z^{\mu\nu}_{g,\mathrm{axial}}\,p_\mu p_\nu \, Q(p)$, where $\mathcal{C}(p)$ is a homogeneous polynomial of degree six in $p$ \cite{Reall:2021voz}. The quadratic factor corresponds to the axial gravitational d.o.f., while $Q(p)$ encodes the characteristic structure of the scalar-polar gravitational sector. Since we are interested in kinetic mixing between the scalar field and the gravitational d.o.f., we henceforth focus on $Q(p)$, which is generically a homogeneous quartic polynomial in $p$. We will sometimes refer to $Q(p)$ as the reduced characteristic polynomial of the mixed system or interchangeably as the multi-d.o.f. dispersion relation.\\
Focusing on the polar sector of the metric perturbations, we adopt the amplitude ansatz \footnote{Our analysis employs the standard Regge-Wheeler gauge, but with $K$ relabeled as $H_3$.}
\begin{align}\label{h-ansatz-polar}
\mathcal{A}^h_{\mu\nu}
=
\begin{pmatrix}
h(r) H_{0} & H_{1} & 0 & 0 \\
H_{1} & H_{2}/f(r) & 0 & 0 \\
0 & 0 & r^{2}\,H_3 & 0 \\
0 & 0 & 0 & r^{2}\sin^{2}\theta\,H_3
\end{pmatrix}
Y_{\ell m} \; ,
\end{align}
where $(t,r,\theta,\varphi)$ are the background coordinates and
$Y_{\ell m}(\theta,\varphi)$ denote the spherical harmonics encoding the
angular dependence of the perturbations. We substitute the eikonal ansatz
into the linearized equations of motion and keep only the leading
$\mathcal{O}(\epsilon^{-2})$ terms.
At this order, only derivatives acting on the phase $S$ contribute.
Consequently, the amplitudes of the metric and scalar perturbations, as
well as the spherical harmonics $Y_{\ell m}$, may be treated as slowly
varying and effectively constant. The field equations therefore reduce to terms
proportional to algebraic, quadratic combinations of $p_\mu$, which define the
dispersion relations for scalar and metric perturbations on the given background.\\

The components of the linearized metric and scalar field equations in the eikonal approximation are presented in Appendix \ref{app:FieldEqsLO}. Although there are initially eleven equations, only two are independent. The redundant components can be used to eliminate all variables except $\mathcal{A}^\phi$ and $H_3$ from ${}^{\phi}\mathcal{E}^{\text{lin}}_{\,|_{\epsilon\rightarrow 0}}$ and ${}^{g}\mathcal{E}^{\text{lin}}_{00\,|_{\epsilon\rightarrow 0}}$. This procedure agrees with the approach described in Appendix B of Ref.~\cite{Blazquez-Salcedo:2016enn}, hence confirming that we are dealing with a mixed two d.o.f. system, but we point out the appearance of genuinely new terms due to the presence of $q$.

These equations can be assembled into a $2\times 2$ matrix $Q^{IJ}(p)$ in
field space, which defines the system of characteristic equations
governing the coupled scalar-metric perturbations. Explicitly, the characteristic system reads
\begin{align}
Q^{IJ}(p)\, t_J = 0
\quad \iff \quad
\begin{pmatrix}
Q^{\phi\phi} & Q^{\phi g} \\
Q^{g\phi} & Q^{gg}
\end{pmatrix}
\begin{pmatrix}
\mathcal{A}^{\phi} \\
H_3
\end{pmatrix}
=
\begin{pmatrix}
0 \\
0
\end{pmatrix},
\label{eq:charSys}
\end{align}
where $I,J$ are field-space indices, with $I=\phi$ corresponding to the scalar perturbation amplitude $\mathcal{A}^\phi$ and $I=g$ to the metric perturbation amplitude $H_3$. The explicit expressions for the components $Q^{IJ}$ are given in Appendix \ref{app:DemixedEqsLO}.

A nontrivial solution exists if and only if the determinant of this
matrix vanishes. Accordingly, one defines
\begin{equation}
Q(p) \equiv \det Q^{IJ}(p),
\end{equation}
which is the aforementioned homogeneous quartic polynomial in the momenta $p$. The
equation $Q(p)=0$ therefore defines the multi-d.o.f. dispersion relation for
the mixed scalar-gravity system. On a generic background this quartic
polynomial does not factorize, reflecting the fact that the scalar and
metric degrees of freedom cannot be fully decoupled; equivalently, the
characteristic surfaces cannot in general be written as the product of
independent acoustic metrics/cones for the two sectors \cite{Reall:2021voz} (see also
Ref.~\cite{Kovacs:2020ywu}).

Being a scalar, $Q$ is independent of the choice of coordinates and provides a convenient
invariant object for the analysis of both stability and screening
properties of the theory. In what follows, we will therefore use $Q$ as the primary diagnostic of the characteristic structure of the coupled system.

We can also view $Q$ as the fully symmetrized contraction of a rank-$4$
tensor with four momenta,
\begin{align}
    Q &\equiv Q^{\mu\nu\rho\sigma} p_\mu p_\nu p_\rho p_\sigma 
    \label{eq:Qtensor-def}
\end{align}
where $Q^{\mu\nu\rho\sigma}$ is totally symmetric under permutations of
its indices.

Because of spherical symmetry, the indices take values
$\mu \in \{0,1,\Omega\}$, where $\Omega$ denotes any angular direction $2,3$. Here the labels $0$,$1$,$2$, and $3$ stand for example for the static coordinates $t$, $r$, $\theta$, and $\varphi$. The tensor components are understood to satisfy
\begin{equation}
Q^{\mu\nu\rho\sigma}
=
Q^{(\mu\nu\rho\sigma)} ,
\qquad
Q^{0022}=Q^{0033},\;
Q^{1122}=Q^{1133},\;
Q^{2222}=Q^{3333},
\end{equation}
with $p_\Omega^2 \equiv p_2^2 + \frac{1}{\sin^2(\theta)}p_3^2$. Equivalently, the sum \eqref{eq:Qtensor-def} may be written in a
combinatorially explicit form as
\begin{equation}
Q
=
\sum_{\mu\le\nu\le\rho\le\sigma}
\frac{4!}{n_\mu!\,n_\nu!\,n_\rho!\,n_\sigma!}\,
Q^{\mu\nu\rho\sigma}\,
p_\mu p_\nu p_\rho p_\sigma ,
\label{eq:Qtensor-sym}
\end{equation}
where $n_\alpha$ counts the number of occurrences of the index
$\alpha\in\{0,1,\Omega\}$ in the ordered set
$\{\mu,\nu,\rho,\sigma\}$, hence ensuring full index symmetrization, while assuming isotropy in the angular directions.

Expanding Eq.~\eqref{eq:Qtensor-sym} explicitly reproduces
\begin{align}
Q &= Q^{0000} p_0^4
   + 4 Q^{0001} p_0^3 p_1
   + 6 Q^{0011} p_0^2 p_1^2
   + 6 Q^{0022} p_0^2 p_\Omega^2 \notag \\
  &\quad
   + 4 Q^{0111} p_0 p_1^3
   + 12 Q^{0122} p_0 p_1 p_\Omega^2
   + Q^{1111} p_1^4 \notag \\
  &\quad
   + 6 Q^{1122} p_1^2 p_\Omega^2
   + Q^{2222} p_\Omega^4 ,
\label{eq:Qcomps-def}
\end{align}

Moreover, it will be useful to define the unique scalar contraction of $Q^{\mu\nu\rho\sigma}$ with the background metric, i.e.
\begin{align}
\text{Tr}\,Q &\equiv Q^{\mu\nu\rho\sigma} g_{\mu\nu} g_{\rho\sigma} = h^2 Q^{0000} - 2 \frac{h}{f} Q^{0011} 
               + \frac{Q^{1111}}{f^2} - 4 h r^2 Q^{0022}  + 4 \frac{r^2}{f} Q^{1122} + \frac{8}{3} r^4 Q^{2222} .
\label{eq:TrQ-def}
\end{align} 
This quantity - which again is a scalar - gives a coordinate-invariant diagnostic of the reduced characteristic polynomial. We can use Eq. \eqref{eq:TrQ-def} to define a central quantity of this paper, the characteristic invariant
\begin{align} \label{screening-I}
\boxed{\mathcal{I} \equiv \frac{\text{Tr}\,Q}{8}}\,,
\end{align}
where the factor $1/8$ is introduced as a normalization relative to the
background metric $g^{\mu\nu}$.\\
As we will discuss in the next section, this structure factorizes into a product of acoustic metrics for each d.o.f. only in simple setups. Such factorization has important implications for assessing both the stability of solutions and the magnitude of screening effects. In general, however, this factorization does not occur, and the multi-d.o.f. dispersion relation cannot be fully decoupled.

Notice that in the axial sector, due to the absence of mixing between the scalar and gravity d.o.f., the analysis involves only the three metric equations from the outset. These can then be straightforwardly reduced to a single characteristic equation, quadratic in the momenta $p$, from which the corresponding acoustic metric $Z^{\mu\nu}_{g,\mathrm{axial}}$ can be read off directly.
We verified that in this case, all the criteria for no ghost instabilities (via signature of acoustic metric), no gradient instabilities (via sign of acoustic metric determinant) as well as the possibility to choose a good Cauchy surface are satisfied for the solutions presented in Sec. \ref{sec:example}.

\subsection{Limiting cases}
In the following, we illustrate - by means of two concrete limiting cases - that $Q$ indeed factorizes in sufficiently simple setups, thereby motivating our choice of the characteristic invariant.
We begin by considering the $K$-essence limit, which corresponds to taking $\alpha \to 0$.
In this limit, the GB term is absent and kinetic mixing vanishes. As a result, the principal symbol becomes the product of the acoustic metric for $K$-essence $Z^{\mu\nu}_\phi$ (giving a quadratic structure in $p$) with the acoustic metric for the polar gravitational sector which reduces to the background metric, i.e. $Z^{\mu\nu}_{g,\,\text{polar}}=g^{\mu\nu}$.\\ 
$Z^{\mu\nu}_\phi$in its general, covariant form reads
\begin{equation}\label{Zmunu-K-essence}
    Z^{\mu\nu}_\phi = K_{X} \, g^{\mu\nu} - K_{XX} \nabla^\mu \phi \nabla^\nu \phi \, .
\end{equation}
Adopting the standard static, spherically symmetric ansatz given in Eq.~\eqref{metric-ansatz}, the non-vanishing components of $Z^{\mu\nu}_\phi$ are
\begin{align}
Z^{tt}_\phi &= -\frac{K_{X} h + K_{XX} q^{2}}{h^{2}}, \\[6pt]
Z^{rt}_\phi &= \frac{K_{XX} f q \,\varphi'}{h}, \\[6pt]
Z^{rr}_\phi &= f \left( K_{X} - K_{XX} f (\varphi')^{2} \right), \\[6pt]
Z^{\theta\theta}_\phi &= \frac{K_{X}}{r^{2}} = Z^{\varphi\varphi}_\phi\sin^{-2}\theta ,
\end{align}
and its trace evaluates to 
\begin{equation}
\text{Tr}\,Z_\phi= Z^{\mu\rho}_\phi g_{\rho\mu}
= - h Z^{tt}_\phi
+ \frac{Z^{rr}_\phi}{f}
+ 2 r^{2} Z^{\theta\theta}_\phi,
\end{equation}
In the limit $\alpha\rightarrow0$, we find that - as expected - that $\text{Tr}Z_\phi$ matches $4\mathcal{I}$, with both expressions yielding
\begin{equation}
 \text{Tr}Z_\phi=   4\,K_{X}
- f\,K_{XX}\,(\varphi')^{2}
+ \frac{K_{XX}\,q^{2}}{h} = \left(K_{X}+2XK_{XX}\right)\left(1+3\frac{K_{X}}{K_{X}+2XK_{XX}}\right)=2\Omega_S\left(1+3 c_s^2\right),
\end{equation}
where we identified the conformal factor $\Omega_S=(K_{X}+2XK_{XX})/2$, as well as the scalar sound speed $c_s^2=\frac{K_{X}}{K_{X}+2XK_{XX}}$, while for the gravitational sector we trivially have $\Omega_T=1$ as well as $c_T=1$. Since the conformal factor $\Omega_S$ encodes both the presence/absence of ghost instabilities and the strength of screening, this result makes explicit that the same information is captured by our characteristic invariant $\mathcal{I}$.\\ 

Next we can specialize to an FLRW background by adopting static coordinates of the de Sitter metric, for which $f =h = 1 - H^2 r_s^2$, with accordingly transformed momentum components, and imposing the cosmological scalar equation of motion Eq. \eqref{cosmo-eq}. This yields explicit expressions for the cosmological reduced characteristic polynomial $Q_c$ and its invariant trace $\mathrm{Tr}\,Q_c$.
For this homogeneous cosmology case (which however features non-zero $\alpha$), we obtain by evaluating Eq. \eqref{eq:TrQ-def}
\begin{align}
\text{Tr}\,Q_c=2 \Bigl(&
K_{XX}\, q_c^2
+ 20\, K_{XX}\, q_c^3 \alpha H
+ 128\, K_{XX}\, q_c^4 \alpha^2 H^2 
+ \frac{32 \bigl( \alpha + 8 K_{XX}\, q_c^6 \alpha^3 \bigr) H^3}{q_c} \notag \\
&+ 960\, \alpha^2 H^4
+ 7680\, q_c \alpha^3 H^5
+ 11264\, q_c^2 \alpha^4 H^6
\Bigr) .
\end{align}
The explicit form of $Q_c$ resulting from Eq. \eqref{eq:Qcomps-def} is not shown here to maintain clarity of presentation.\\
On the other hand, using the commonly used EFT parameterization for general scalar-tensor models, it is known from the quadratic action for perturbations on the homogeneous cosmological background, e.g. \cite{Bellini:2014fua}, that - due to the symmetry of the background - factorization of the principal symbol into two acoustic metrics also occurs here. One finds in the Friedmann coordinates

\begin{align}
Z^{\mu\nu}_\phi  &= 2\, \Omega_S \, 
\mathrm{diag}\bigl(-1,\, c_s^2,\, c_s^2,\, c_s^2\bigr), \\[2mm]
Z^{\mu\nu}_g &= 2\, \Omega_T \, 
\mathrm{diag}\bigl(-1,\, c_T^2,\, c_T^2,\, c_T^2\bigr) .
\end{align}
Here we used the definitions $\Omega_S
= \frac{2 \mathcal{D} M_*^{2}}{(2 - \alpha_B)^{2}}$ and $\Omega_T = \frac{M_*^{2}}{8}$ for the two conformal factors, as well as $c_s^{2}
= - \frac{(2 - \alpha_B)}{H^{2} \mathcal{D}}  
\left[
\mathcal{H}
- \frac{1}{2} \alpha_B (1 + \alpha_T) H^{2}
\right]$ and $c_T^2=1 + \alpha_T$ for the cosmological sound-speeds of scalar and tensor fluctuations respectively. Moreover, we used the explicit expressions $M_*^{2}= 1 + 8 \alpha\, q_c H$ and $\mathcal{D}
= \alpha_K + \frac{3}{2}\alpha_B^{2}$, where $\alpha_K
= \frac{
q_c^{2} \left( K_X + q_c^{2} K_{XX} \right)
- 8 \alpha\, q_c H^{3}
}{
H^{2} M_*^{2}
}$, $\alpha_T = - \frac{8 \alpha\, q_c H}{M_*^{2}}=\alpha_B$, as well as $\mathcal{H} = H^{2} \alpha_T$.
This once more underscores the importance of the two conformal factors appearing in front of both acoustic metrics, especially when assessing the presence of ghost instabilities. At the tensorial level this implies
\begin{equation}
Q^{\mu\nu\rho\sigma}_{\text{EFT}}
=
Z^{(\mu\nu}_\phi Z^{\rho\sigma)}_g ,
\end{equation}
and hence 
\begin{equation}
Q_{\text{EFT}}(p)
=
Z^{(\mu\nu}_\phi Z^{\rho\sigma)}_g p_\mu p_\nu p_\rho p_\sigma ,
\label{eq:EFTQ}
\end{equation}
From symmetrization, the trace of Eq. \eqref{eq:EFTQ} evaluates to
\begin{equation}
\mathrm{Tr}\,Q_{\text{EFT}}
=    Z^{(\mu\nu}_\phi Z^{\rho\sigma)}_g
g_{\mu\nu} g_{\rho\sigma}
=
\frac{1}{3}
\left(
\text{Tr}\,Z_\phi\text{Tr}\,Z_g
+ 2\,\mathrm{Tr}(Z^{\mu\rho}_\phi Z^{\;\;\;\nu}_{g, \rho})
\right)=4\Omega_S\Omega_T\left(1+c_s^2+c_T^2+5c_s^2 c_T^2\right).
\end{equation}

This identity again is purely algebraic and does not depend on coordinates. A direct comparison between the cosmological and EFT quantities shows that they coincide once a gauge factor $\Gamma\equiv-\,\frac{q_c}{H\, M_*^{2}\,(2 - \alpha_B)}$ is introduced to account for the choice of unitary gauge adopted in Ref.~\cite{Bellini:2014fua}. More precisely, one finds $Q_{\text{EFT}}(p)=\Gamma^2Q_c(p) $, as well as $\text{Tr}\,Q_{\text{EFT}}(p)=\Gamma^2 \text{Tr}\,Q_c(p)$ where, in the second relation, $K_X$ has to be fixed using the cosmological scalar Eq.~\eqref{cosmo-eq}.\\

These two examples demonstrate that, in the corresponding limits where the mixed scalar-gravity system factorizes, the characteristic invariant reduces to the product of the two acoustic metrics associated with the individual d.o.f. This leads us to highlight three key observations that motivate the definition of the characteristic invariant $\mathcal{I}$ in Eq.~\eqref{screening-I}, and why it can - among other things - be used as a screening indicator:
\begin{itemize}
    \item By construction, cf.~Eq.~\eqref{eq:TrQ-def}, $\mathcal{I}$ is a scalar quantity and is therefore coordinate invariant.
    \item The invariant  encodes the product of the conformal factors associated with all acoustic metrics. Consequently, its sign allows one to infer the proto-stability of the system, namely whether one d.o.f is ghostly relative to the other. This provides the best available diagnostic in situations where the individual conformal factors cannot be disentangled, as is generically the case. Naturally, if both d.o.f. are ghostly, this would only become apparent upon comparison with an additional healthy d.o.f. (for instance, in the axial sector).
    \item Owing again to the presence of the conformal factors, the magnitude of $\mathcal{I}$ offers an estimate of the strength of the screening effect, corresponding to invariant eigenvalues of the acoustic metrics becoming large. Conversely, a very small value of $\mathcal{I}$ signals the onset of a strong-coupling regime.
\end{itemize}

\section{Explicit example}\label{sec:example}

In this section we will now discuss a concrete example of a theory of the form \eqref{action}, where it is possible to find both self-accelerating cosmological solutions and black-hole ones. While only a toy model, it will serve us as an analytically tractable example to study stability and screening using the approach introduced in the previous section.

For concreteness let us consider the theory \eqref{action} with the kinetic term $K(X) = \eta X + c_2 X^2/M^2$, with $\eta = \pm 1$,
\begin{equation}
    S = M_P^2 \int d^4 \! x \sqrt{-g} \left[ \frac{R}{2} - \frac{\eta}{2} (\partial \phi)^2 + \frac{c_2}{4M^2} (\partial \phi)^4 + \alpha \phi \, \mathcal{G} \right] \, .
\end{equation}
This is similar to the example discussed in Sec.~\ref{sec:flat} concerning asymptotically flat black holes, but with the remarkable difference that the terms in $K(X)$ will have different signs, namely $\eta < 0$ and $c_2 > 0$. As we will see, this is required for self-accelerating solutions.

\subsection{Cosmology}

For this specific choice of $K(X)$ above, the $J_0=0$ condition defining the attractor solution, Eq.~\eqref{cosmo-eq}, now reads
\begin{eqnarray} \label{cosmo-eq-ex}
    q_c \left( \eta + \frac{c_2 q_c^2}{M^2} \right) = 8 \alpha H^3 \, ,
\end{eqnarray}
which admits either $q_c = \mathcal{O}(\alpha H^3)$, or 
\begin{eqnarray} \label{cosmo-qc-ex}
    q_c^2 = - \frac{\eta}{c_2} M^2 + \mathcal{O}(\alpha H^3 M) \, .
\end{eqnarray}
The latter requires $\eta c_2 < 0$. On the other hand, the Friedmann equation with no cosmological constant reads,
\begin{eqnarray} \label{Friedmann-eq-ex}
    3 H^2 = \frac{\eta q_c^2}{2} + \frac{3}{4} \frac{c_2 q_c^4}{M^2} - 24 \alpha H^3 q_c \, ,
\end{eqnarray}
which we can evaluate for the above solutions to fix $H$. For the $q_c = \mathcal{O}(\alpha H^3)$ branch the only acceptable solution under our assumption $\sqrt{\alpha} \ll H^{-1}$ is $H = 0$, i.e. Minkowski. Instead, for the nontrivial branches \eqref{cosmo-qc-ex} we have
\begin{eqnarray} \label{self-acc-sol}
    12 c_2 H^2 = M^2 + \mathcal{O}(\alpha H^3 M) \, ,
\end{eqnarray}
where we used $\eta^2 = 1$. Here we find that $c_2 > 0$ is required, unlike the asymptotically flat case\footnote{However, standard flat-space positivity bounds \cite{Adams:2006sv} do not directly apply in this case.}, and therefore from the previous condition we get $\eta < 0$. The latter means that the Minkowski branch ($H = 0$, $q_c = \mathcal{O}(\alpha H^3)$) is unstable, leaving only the self-accelerating solutions that we were looking for, with
\begin{eqnarray}
    q_c^2 = \frac{H^2}{12} + \mathcal{O}(\alpha H^4) \, .
\end{eqnarray}
Notice that if we instead wanted to have a stable Minkowski vacuum, we would need $\eta > 0$ which forbids the self-accelerating solution, i.e. they are mutually exclusive.

\subsection{Black holes}

We now turn to consider static black holes. With the above choice of $K(X)$, the scalar field equation \eqref{scalar-eom} corresponding to the ans\"atze \eqref{metric-ansatz} and \eqref{scalar-ansatz} takes the form of a depressed cubic equation for $\varphi'$, i.e. with no quadratic term,
\begin{eqnarray} \label{scalar-eom-ex}
     \frac{c_2}{M^2} f \left(\varphi'\right)^3 - \left( \eta + \frac{c_2}{M^2} \frac{q^2}{h} \right) \varphi' + \left[ \frac{4\alpha h'}{r^2 h}(1-f) - \frac{\alpha_\text{BH} r_s}{r^2} \frac{1}{\sqrt{h f}} \right] = 0 \, ,
\end{eqnarray}
where a priori both $q$ and $\alpha_\text{BH}$ are the free parameters that determine the solution. The relevant solution must be real for all $r$, at least in the region between the two gravitational horizons, be regular (in the sense of Eq.~\eqref{near-horizon-varphip}), exhibit the correct asymptotics, and furthermore be stable under perturbations. All of these requirements, may or may not impose restrictions on the choice of parameters $q$ and $\alpha_\text{BH}$. 

\begin{figure}[h]
    \centering
    \includegraphics[width=2.7in]{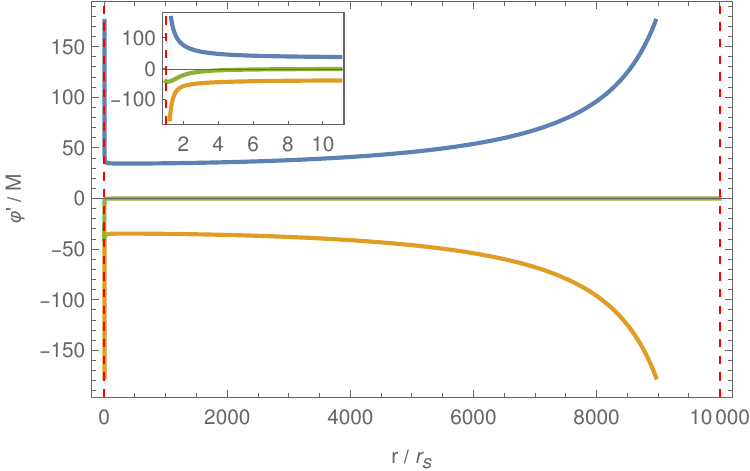}
    \caption{Three branches of solutions to the scalar field equation \eqref{scalar-eom-ex}, for $H r_s = 10^{-4}$, $\alpha_\text{BH} = 2$, and $q = 10 q_c$. For these choices all three branches are real, but only the lower branch ($\varphi' < 0$, orange) satisfies the correct near-horizon form, Eq.~\eqref{near-horizon-varphip}. Inset: Zoom-in to the vicinity of the black-hole horizon.
    }
    \label{fig:branches}
\end{figure}

In order to proceed in finding analytic solutions and analyzing these aspects, for the rest of this paper we will work in a test-field approximation by ignoring backreaction and fixing the background metric functions to Schwarzschild-de Sitter $f = h = 1 - r_s/r - H^2 r^2$, with $H r_s \ll 1$. The solutions for $\varphi'(r)$ may be immediately obtained for all three branches (see Fig.~\ref{fig:branches}). The first aspect that becomes immediately clear is that the near-horizon form \eqref{near-horizon-varphip}, needed to make both the full $\phi$ and its kinetic term $X$ finite at the cosmological horizon, singles out the one branch with $\varphi' < 0$ everywhere. Importantly, this branch may be real or complex depending on the choice of $q$ and $\alpha_\text{BH}$. With the analytic solution at hand we may next then evaluate the characteristic invariant \eqref{screening-I} and use it to establish a necessary condition for stability of perturbations. As discussed in the previous section, $\mathcal{I} < 0$ is associated with ghost instabilities.

\begin{figure}[h]
    \centering
    \hspace{0.3in}
    \includegraphics[width=3in]{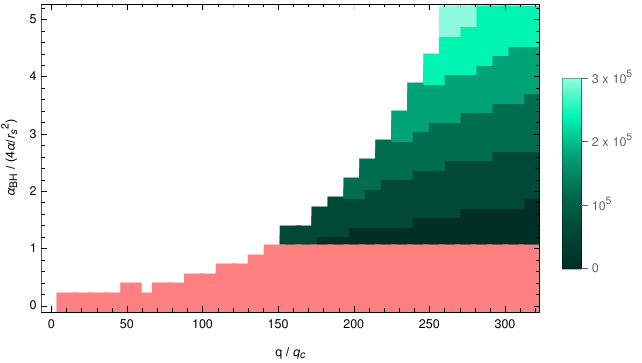}
    \caption{Region in the parameter space of the solutions to the scalar field equation \eqref{scalar-eom-ex} for $H r_s = 10^{-8}$, indicating where the regular branch with correct cosmological asymptotics is real (in color). The plot extends symmetrically to $q<0$. The proto-stable ($\mathcal{I} > 0$) and unstable ($\mathcal{I} < 0$) subregions are colored green and red, respectively. Within the former, different shades of green indicate the changing peak value of $\mathcal{I}$, which is seen to increase towards larger values of $\alpha_\text{BH}$  as expected, with $\mathcal{I} \gg 1$ signaling the presence of kinetic screening. Good solutions exist for continuous values of the scalar charge $ > 4\alpha/r_s^2$, making it a \emph{primary} charge.}
    \label{fig:alphaBH-q}
\end{figure}

Both aspects, reality and stability, end up being important in this specific model to restrict the valid choices of parameters $q$ and $\alpha_\text{BH}$. The regions of proto-stable $\mathcal{I} > 0$ and unstable $\mathcal{I} < 0$ real solutions in the $q$-$\alpha_\text{BH}$ plane are shown in Fig.~\ref{fig:alphaBH-q}. After applying these criteria we observe that there is still an open region of parameter space where, as far as our analysis goes, solutions are valid. This has two consequences that are worth pointing out:
\begin{itemize}
    \item The scalar charge $\alpha_\text{BH}$ is \emph{primary}: It can take continuous values. These are restricted to be strictly larger than the standard scalar-Gauss-Bonnet charge $4\alpha/r_s^2$ to avoid strong coupling ($|\mathcal{I}| \ll 1$) and ghost-like instabilities ($\mathcal{I} < 0$).
    \item The \emph{local} time-derivative $q$ is bounded from below to ensure the existence of real solutions. This bound increases for larger $\alpha_\text{BH}$.
\end{itemize}

The above analysis confirms that the presence of a non-canonical kinetic term $K(X)$ in the theory relaxes the regularity conditions \eqref{reg-cond-BH} and \eqref{reg-cond-cosmo} that require the scalar charge to have a fixed value, i.e. to be \emph{secondary}, promoting it to a free parameter of the solutions, that is, a \emph{primary} charge. 

The second aspect is that $q=0$ is not allowed, and in fact there is a critical value $q_\text{crit}$, such that the solution is real \emph{and} proto-stable only when $|q| \geq q_\text{crit}$. It is interesting to look into detail at the scaling of $q_\text{crit}$ with the separation of scales $H r_s$. For this we first notice by numerical inspection that the good branch ($\varphi' < 0$) is one of the two that may turn complex\footnote{The other possibility being that it is the one that remains real.}, and therefore we may assess whether it is real by looking at the sign of the discriminant,
\begin{eqnarray}
    \Delta(r) = \frac{27 c_2^3 f^9}{M^{12} r^4 h^3} &\Biggl(& 4 c_2^3 q^6 r^4+12 c_2^2 \eta 
   M^2 q^4 r^4 h - 432 \alpha ^2 c_2 M^4 f^3 h h'^2+864 \alpha
   ^2 c_2 M^4 f^2 h h'^2 \nonumber \\ 
   &&-432 \alpha ^2 c_2 M^4 f h
   h'^2+216 \alpha \alpha_\text{BH} c_2 M^4 r_s
   \sqrt{f} (1 - f) h^{3/2} h' \nonumber \\
   &&+12 c_2 M^4 q^2 r^4 h^2-27
   \alpha_\text{BH}^2 c_2 M^4 r_s^2 h^2+4 \eta M^6 r^4
   h^3 \Biggr) \, .
\end{eqnarray}
The reality condition $\Delta(r) \geq 0$ must be imposed for all $r$ of interest. We may express this condition in a simple form by expanding perturbatively on the large separation of scales $H r_s \ll 1$. After some analysis, we obtain the condition
\begin{eqnarray} \label{qcrit}
    \frac{q_\text{crit}}{q_c} \simeq \frac{1}{\sqrt{3}} \left( \frac{\alpha_\text{BH} \sqrt{2 c_2}}{H r_s} \right)^{1/3} \, .
\end{eqnarray}
As the separation of scales becomes larger --- smaller $H r_s$ --- this critical value increases. For realistic values of $H r_s \sim 10^{-21}$, and $\alpha_\text{BH} \sim \mathcal{O}(1)$, this becomes $q_\text{crit} \sim 10^7 \, q_c$. Following the discussion surrounding Eq.~\eqref{approach-to-homogeneity}, this would make a steady-state solution have a very slow approach to homogeneity way outside of the Hubble horizon. In fact, such a steady-state solution is not actually consistent for a $q \gtrsim q_\text{crit}$ and $\alpha_\text{BH} \geq 4\alpha/r_s^2$, as the accretion rate associated to the energy-flux \eqref{energy-flux} is much larger than $H$. Indeed, the corresponding rate of change of the black-hole mass is of order $\dot{r}_s/r_s \sim 10^8 H$ --- much faster than $H$ but much slower than $r_s^{-1}$. The solutions with a static background metric \eqref{metric-ansatz} may then still be considered as quasi-stationary from the point of view of the astrophysical scales.

\begin{figure}[h]
    \centering
    \includegraphics[width=2.7in]{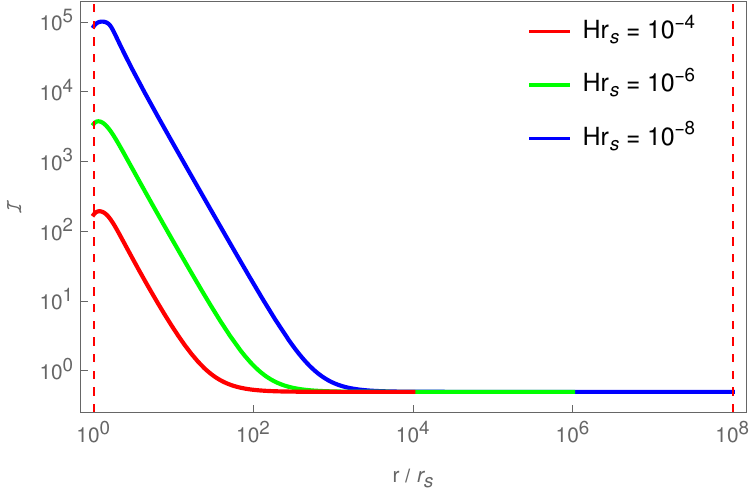}
    \caption{Value of $\mathcal{I}$ for example test-field solutions to the scalar field equation on a Schwarzschild-de Sitter background metric, with $H r_s \sim 10^{-4}, 10^{-6}, 10^{-8}$, in red, green, and blue, respectively. The scalar charge is fixed at $\alpha_\text{BH} = 2$, while the time-gradient $q$ is adjusted in each case to be mildly larger than $q_\text{crit}$ [Eq.~\eqref{qcrit}], i.e. $q=10 q_c,50 q_c,200 q_c$. The quantity $\mathcal{I}$ is closely related to the conformal factor $\Omega_S$ of the scalar-like perturbations, serving as a measure of screening. The effect can be observed to be active, i.e. $\mathcal{I} \gg 1$, within a region of size $r_k$ surrounding the black hole, which grows with the separation of scales. The peak strength $\mathcal{I}_\text{max}$ also scales similarly, but it is always located in close proximity to the black-hole horizon. The value of $\mathcal{I}$ remains large at the horizon proper.}
    \label{fig:screening}
\end{figure}

The final aspect that we wish to study is the kinetic screening. In the previous section we examined two limiting cases where the scalar and gravitational perturbations decouple, showing that the absolute value of $\mathcal{I}$ reduces to the product of the two conformal factors $\Omega_S$ and $\Omega_T$ of each individual acoustic metric for each mode. These factors encode, among other things, information about the screening --- a large $\Omega_S$ and/or $\Omega_T$ gives a large $\mathcal{I}$. We plot $\mathcal{I}$ for an example solution in Fig.~\ref{fig:screening}, confirming the expected increase of several orders of magnitude in the vicinity of the black-hole horizon. The dependence of the peak value of $\mathcal{I}$ with the solution parameters can instead be appreciated in Fig.~\ref{fig:alphaBH-q}, which shows an approximately linear increase with $\alpha_\text{BH}$. A large $\mathcal{I}$ reflects the growth of either or both of the conformal factors $\Omega_S$ and $\Omega_T$. 
While the information is not sufficient to pinpoint which one is growing, based on the assumption of perturbativity in $\alpha$ we may conclude that $\Omega_S \gg 1$, and $\Omega_T \sim 1$. In such case it is numerically accurate to study the modes separately, as if there was no kinetic mixing. In this case, the acoustic metric for the scalar mode is basically that of $K$-essence \eqref{Zmunu-K-essence}, and the corresponding conformal factor $\Omega_S = (K_X + 2X K_{XX})/2$, as described in the previous section. Nevertheless, the more general treatment through the computation of $\mathcal{I}$ allows to explore beyond this perturbative regime to assess the presence of screening.

\section{Conclusions}\label{sec:conclusion}

Black holes provide a very interesting laboratory for testing General Relativity and its possible extensions thanks to the detection of the gravitational waves that are produced during the late inspiral and merger phases of BH coalescence. This prospect becomes particularly tantalizing when the modification of gravity may be responsible for the accelerated expansion of the universe ---the elusive Dark Energy. A common misconception states that the physics at cosmological scales ---the deep infrared--- cannot affect the shorter-scale physics in the strong-field regime surrounding black holes. To the contrary, Effective Field Theories of modified gravity inevitably contain derivative interactions, which tend to become increasingly important at shorter scales. In a wide class of models contained in shift-symmetric scalar-tensor theories, these derivative interactions are in fact primarily responsible for the acceleration, making them not only important but dominant at shorter scales. In the presence of a nontrivial background for the scalar field that breaks Lorentz invariance spontaneously, the propagation of perturbations is strongly affected by such higher-dimensional operators. This has a number of consequences that are very relevant for the phenomenology of black holes and their gravitational wave signals, including a modified sound speed, and a potentially large suppression of scalar interactions ---a screening mechanism. While all these aspects have been studied separately, there have been few attempts to make it all fit together.

In this paper we have approached the construction of hairy black-hole solutions in the simplest model where: \emph{i)} self-accelerating cosmological solutions exist, \emph{ii)} black holes carry a scalar charge. These requirements led to our choice of a $K$-essence type theory, with the addition of a scalar-Gauss-Bonnet coupling \eqref{action}. We initially considered asymptotically-flat black holes, allowing a comparison to the treatment of NS usually done in the literature. Then in the full setup we considered black-holes embedded into a self-accelerating cosmology. By studying the regularity conditions at the black-hole and cosmological horizons, we found that the scalar charge $\alpha_\text{BH}$ can \emph{a priori} take continuous values, i.e. a \emph{primary} charge, owing to both the nontrivial kinetic term $K(X)$ and the non-vanishing scalar time-derivative $q$. This contrasts sharply with the fixed \emph{secondary} scalar-Gauss-Bonnet charge for a static, asymptotically-flat configuration, or when the kinetic term is canonical. This is a novel finding of this paper.

In the context of a specific example, we confirmed analytically in a test-field approximation that the scalar charge remains of \emph{primary} type, even though is subject to some restrictions from the existence of real, regular solutions with the correct near-horizon behaviour, in addition to a stability requirement. The latter was posed in term of a characteristic invariant that is computed from the multi-degree-of-freedom dispersion relation, in this case of quartic order, necessary to deal with the kinetic mixing induced by the scalar-Gauss-Bonnet operator. This is a second-best quantity to the conformal factor of the acoustic metric for perturbations, which holds information about ghost instabilities, strong-coupling and screening, but it is only directly accessible by the full decoupling of the modes. 

Armed with the characteristic invariant we were also able to assess the presence of kinetic screening in the vicinity of the black hole. The next step in the direction of quantifying observable effects would be to compute the rate of energy loss due to scalar-wave emission during the inspiral phase of the merger. On the one hand, this is an important effect that contributes to the dephasing of the GW signal with respect to the GR one, making it one of the strongest ways to constrain a scalar-Gauss-Bonnet coupling in the absence of screening. On the other hand, numerical studies of kinetic screening for binary neutron stars \cite{Bezares:2021dma, Cayuso:2024ppe} suggest that while dipolar scalar radiation is suppressed, the mechanism becomes less efficient for quadrupolar scalar radiation in dynamical situations. The setup considered here, however, differs in several aspects to the one studied, as it concerns black-holes with nontrivial regularity conditions, it has a background scalar configuration with a time-like gradient, and features kinetic mixing between scalar and gravitational perturbations. Therefore, a direct translation of the results of Ref.~\cite{Bezares:2021dma, Cayuso:2024ppe} is not possible. In this context, an analytic computation of the emitted power faces severe challenges due to the nonlinearities of the background solutions, especially for a dynamical, non-symmetric two-body scenario. One possible approach is to perturb around the spherically-symmetric screened profile and study the modified propagation on top, whose details dictates the efficiency of the energy loss \cite{Dar:2018dra}. This approach becomes more involved by the impossibility of completely decoupling the modes of the linear perturbations, as in our model. To aid in this type of computation, it would be interesting to consider the construction of other, independent, characteristic invariants which would allow to extract even more information about the individual acoustic metrics of each mode. Alternatively, an approach to the computation of the emitted power directly from the full two-degree-of-freedom system may be devised.

An interesting consequence of screening around BHs is that it opens up a different range of masses as compared to the usual matter-sourced screening, allowing to go up to supermassive BHs. The intensity of the screening effect depends on the ratio of the screening radius to the size of the source, \(r_k/r_s \propto (\alpha_\text{BH}/m_\text{BH})^{1/2}\) (see Sec.~\ref{sec: screening for bhs asympt flat}), leading to potentially larger beyond-GR effects for more massive objects. This goes against the common belief that low-mass BHs are the place where the strongest constraints may be obtained for a scalar-Gauss-Bonnet type of scalar charge. In the case of a time-dependent scenario there is another novel aspect associated to the scalar charge becoming $\emph{primary}$, but with a lower bound that is basically the corresponding static charge. This may open new scenarios for testing their effects as the scalar charge is no longer uniquely tied to the mass. For example, two equal-mass black holes usually do not emit dipolar scalar radiation, as the scalar charges are considered to be equal~\cite{Yagi:2015oca}. Instead, if equal-mass BHs are allowed to have different scalar charges, this will lead to an enhancement of GR deviations through dipolar emission.

In light of the recent DESI \cite{DESI:2025zgx} results, it is also relevant to generalize to more complicated models beyond the one considered here, especially beyond $K$-essence on the cosmological end of the problem. Some work has been done recently in this direction by considering the cubic Galileon in addition to the scalar-Gauss-Bonnet operator, but only for asymptotically flat spacetimes \cite{Thaalba:2025lwe}. Further work is required to construct explicit hairy black-hole solutions not only for those operators that may be relevant for cosmology, but others that only kick-in at shorter-scales and still play an important role in black-hole physics \cite{Noller:2019chl}, especially with cosmological asymptotics. In all these cases, screening plays a fundamental role in shaping the phenomenology at the crossroads of Dark Energy and black holes.

\begin{acknowledgments}
The authors would like to thank Ignacy Sawicki for helpful discussions. The work of G.T. was supported by the Czech Science Foundation (GAČR) project PreCOG (Grant No.\ 24-10780S) and by IBS under the project code IBS-R018-D3. The work of L.G.T.\ was supported by the European Union (Grant No.\ 101063210) and the European Structural and Investment Funds and the Czech Ministry of Education, Youth and Sports (Project FORTE CZ.02.01.01/00/22 008/0004632). G.L. gratefully acknowledges the Central European Institute of Cosmology (CEICO), Prague, Czech Republic, for their generous hospitality during his visit. The authors would also like to thank the hospitality of the Centro de Ciencias de Benasque Pedro Pascual during the initial stages of this work.
\end{acknowledgments}

\bibliography{bibliography}

\appendix

\section{Components of the linearized field equations in the eikonal limit}\label{app:FieldEqsLO}

The components for the linearized metric field equations in the eikonal limit, ${}^{g}\mathcal{E}^{\text{lin}}_{\mu\nu\,|_{\epsilon\rightarrow 0}}=0$, read
\small
\begin{align}
{}^{g}\mathcal{E}^{\text{lin}}_{01\,|_{\epsilon\rightarrow 0}} &= -\frac{1}{4 h r^3} \Bigl[ 
h r \Bigl( H_1 M_p^2 p_2^2 + H_1 M_p^2 p_3^2 \csc^2\theta 
+ 2 p_0 p_1 (8 \mathcal{A}^\phi \alpha + H_3 M_p^2 r^2) \Bigr) \nonumber\\
&\quad + 4 H_3 q \alpha (p_2^2 + p_3^2 \csc^2\theta) r h' 
- 8 f h \alpha \Bigl( 2 \mathcal{A}^\phi p_0 p_1 r 
+ (H_1 p_2^2 + H_1 p_3^2 \csc^2\theta + 2 H_3 p_0 p_1 r^2) \varphi' \Bigr) \Bigr], \\[0.8em]
{}^{g}\mathcal{E}^{\text{lin}}_{00\,|_{\epsilon\rightarrow 0}} &= \frac{h}{4 r^3} \Bigl[
16 f^2 p_1^2 \alpha r (\mathcal{A}^\phi + H_3 r \varphi') 
- (p_2^2 + p_3^2 \csc^2\theta) \Bigl( (H_2 + H_3) M_p^2 r - 4 \alpha f' (2 \mathcal{A}^\phi + H_3 r \varphi') \Bigr) \nonumber\\
&\quad + f \Bigl( -2 p_1^2 r (8 \mathcal{A}^\phi \alpha + H_3 M_p^2 r^2) 
+ 8 H_2 \alpha (p_2^2 + p_3^2 \csc^2\theta) \varphi' 
+ 8 H_3 \alpha (p_2^2 + p_3^2 \csc^2\theta) r \varphi'' \Bigr) \Bigr], \\[0.8em]
{}^{g}\mathcal{E}^{\text{lin}}_{11\,|_{\epsilon\rightarrow 0}} &= \frac{1}{4 f h r^3} \Bigl[
\frac{1}{2} h (H_0 - H_3) M_p^2 (-p_2^2 - 2 p_3^2 + p_2^2 \cos 2\theta) \csc^2\theta \, r 
- 2 p_0^2 r (8 \mathcal{A}^\phi \alpha + H_3 M_p^2 r^2) \nonumber\\
&\quad + 4 f \alpha \Bigl( 2 h H_0 (p_2^2 + p_3^2 \csc^2\theta) \varphi' 
+ 4 p_0^2 r (\mathcal{A}^\phi + H_3 r \varphi') \nonumber\\
&\qquad + \frac{1}{2} (-p_2^2 - 2 p_3^2 + p_2^2 \cos 2\theta) \csc^2\theta \, h' (2 \mathcal{A}^\phi + H_3 r \varphi') \Bigr) \Bigr], \\[0.8em]
{}^{g}\mathcal{E}^{\text{lin}}_{22\,|_{\epsilon\rightarrow 0}} &= \frac{1}{4} \Biggl[
\frac{4 f \alpha h' (2 q (H_1 p_3^2 \csc^2\theta + H_3 p_0 p_1 r^2) + \mathcal{A}^\phi p_3^2 \csc^2\theta h')}{h^2} 
+ 8 f^2 H_0 p_1^2 \alpha r \varphi' \nonumber\\
&\quad + p_3^2 \csc^2\theta ((-H_0 + H_2) M_p^2 + 4 H_0 \alpha f' \varphi') \nonumber\\
&\quad + f \Bigl( (-H_0 + H_3) M_p^2 p_1^2 r^2 + 8 H_0 p_3^2 \alpha \csc^2\theta \varphi'' \Bigr) \nonumber\\
&\quad + \frac{1}{h} \Bigl( - (H_2 + H_3) M_p^2 p_0^2 r^2 + 4 \alpha f' (-\mathcal{A}^\phi p_3^2 \csc^2\theta h' + p_0^2 r (2 \mathcal{A}^\phi + H_3 r \varphi')) \nonumber\\
&\quad - 4 f^2 p_1 \alpha r (4 H_1 p_0 \varphi' + p_1 h' (2 \mathcal{A}^\phi + H_3 r \varphi')) \nonumber\\
&\quad + 2 f \bigl( 2 H_2 \alpha (2 p_0^2 r - p_3^2 \csc^2\theta h') \varphi' - 4 \mathcal{A}^\phi p_3^2 \alpha \csc^2\theta h'' + p_0 r^2 (H_1 M_p^2 p_1 + 4 H_3 p_0 \alpha \varphi'') \bigr) \Bigr) \Biggr], \\[0.8em]
{}^{g}\mathcal{E}^{\text{lin}}_{33\,|_{\epsilon\rightarrow 0}} &= \frac{\sin^2 \theta}{4 h^2} \Biggl[
4 f \alpha h' (2 q (H_1 p_2^2 + H_3 p_0 p_1 r^2) + \mathcal{A}^\phi p_2^2 h') \nonumber\\
&\quad + h^2 \Bigl( 8 f^2 H_0 p_1^2 \alpha r \varphi' + p_2^2 ((-H_0 + H_2) M_p^2 + 4 H_0 \alpha f' \varphi') \nonumber\\
&\quad + f ((-H_0 + H_3) M_p^2 p_1^2 r^2 + 8 H_0 p_2^2 \alpha \varphi'') \Bigr) \nonumber\\
&\quad - h \Bigl( (H_2 + H_3) M_p^2 p_0^2 r^2 - 4 \alpha f' (-\mathcal{A}^\phi p_2^2 h' + p_0^2 r (2 \mathcal{A}^\phi + H_3 r \varphi')) \nonumber\\
&\quad + 4 f^2 p_1 \alpha r (4 H_1 p_0 \varphi' + p_1 h' (2 \mathcal{A}^\phi + H_3 r \varphi')) \nonumber\\
&\quad - 2 f (2 H_2 \alpha (2 p_0^2 r - p_2^2 h') \varphi' - 4 \mathcal{A}^\phi p_2^2 \alpha h'' + p_0 r^2 (H_1 M_p^2 p_1 + 4 H_3 p_0 \alpha \varphi'')) \Bigr) \Biggr], \\[0.8em]
{}^{g}\mathcal{E}^{\text{lin}}_{12\,|_{\epsilon\rightarrow 0}} &= \frac{p_2}{4 h^2 r} \Bigl[ 4 H_3 p_0 q \alpha r h' + h^2 p_1 ((-H_0 + H_3) M_p^2 r + 8 f H_0 \alpha \varphi') \nonumber\\
&\quad + h (H_1 M_p^2 p_0 r - 4 f \alpha (2 H_1 p_0 \varphi' + p_1 h' (2 \mathcal{A}^\phi + H_3 r \varphi'))) \Bigr], \\[0.8em]
{}^{g}\mathcal{E}^{\text{lin}}_{13\,|_{\epsilon\rightarrow 0}} &= - \frac{p_3}{4 h^2 r} \Bigl[ 4 H_3 p_0 q \alpha r h' + h^2 p_1 ((-H_0 + H_3) M_p^2 r + 8 f H_0 \alpha \varphi') \nonumber\\
&\quad + h (H_1 M_p^2 p_0 r - 4 f \alpha (2 H_1 p_0 \varphi' + p_1 h' (2 \mathcal{A}^\phi + H_3 r \varphi'))) \Bigr], \\[0.8em]
{}^{g}\mathcal{E}^{\text{lin}}_{02\,|_{\epsilon\rightarrow 0}} &= \frac{f H_3 p_1 p_2 q \alpha h'}{h} + \frac{p_2}{4 r} \Bigl[ -(H_2 + H_3) M_p^2 p_0 r - 8 f^2 H_1 p_1 \alpha \varphi' \nonumber\\
&\quad + 4 p_0 \alpha f' (2 \mathcal{A}^\phi + H_3 r \varphi') + f (H_1 M_p^2 p_1 r + 8 H_2 p_0 \alpha \varphi' + 8 H_3 p_0 \alpha r \varphi'') \Bigr], \\[0.8em]
{}^{g}\mathcal{E}^{\text{lin}}_{03\,|_{\epsilon\rightarrow 0}} &= \frac{f H_3 p_1 p_3 q \alpha h'}{h} + \frac{p_3}{4 r} \Bigl[ -(H_2 + H_3) M_p^2 p_0 r - 8 f^2 H_1 p_1 \alpha \varphi' \nonumber\\
&\quad + 4 p_0 \alpha f' (2 \mathcal{A}^\phi + H_3 r \varphi') + f (H_1 M_p^2 p_1 r + 8 H_2 p_0 \alpha \varphi' + 8 H_3 p_0 \alpha r \varphi'') \Bigr], \\[0.8em]
{}^{g}\mathcal{E}^{\text{lin}}_{23\,|_{\epsilon\rightarrow 0}} &= \frac{p_2 p_3}{4 h^2} \Bigl[ -4 f \alpha h' (2 H_1 q + \mathcal{A}^\phi h') + 4 h \alpha (\mathcal{A}^\phi f' h' + f (H_2 h' \varphi' + 2 \mathcal{A}^\phi h'')) \nonumber\\
&\quad + h^2 ((H_0 - H_2) M_p^2 - 4 H_0 \alpha f' \varphi' - 8 f H_0 \alpha \varphi'') \Bigr],  
\end{align}
\normalsize

while for the scalar field equation one obtains
\small
\begin{align}
{}^{\phi}\mathcal{E}^{\text{lin}}_{\,|_{\epsilon\rightarrow 0}}
 &= \frac{4 H_2 p_0^2 \alpha}{h r^2} - \frac{4 f H_2 p_0^2 \alpha}{h r^2} - \frac{8 f H_1 p_0 p_1 \alpha}{h r^2} + \frac{8 f^2 H_1 p_0 p_1 \alpha}{h r^2} \nonumber\\
&\quad + \frac{4 f H_0 p_1^2 \alpha}{r^2} - \frac{4 f^2 H_0 p_1^2 \alpha}{r^2} 
- \frac{\mathcal{A}^\phi p_0^2 K_X}{h} + \mathcal{A}^\phi f p_1^2 K_X \nonumber\\
&\quad + \frac{\mathcal{A}^\phi p_2^2 K_X}{r^2} + \frac{\mathcal{A}^\phi p_3^2 \csc^2\theta K_X}{r^2} - \frac{\mathcal{A}^\phi p_0^2 q^2 K_{XX}}{h^2} \nonumber\\
&\quad - \frac{2 H_0 p_2^2 \alpha f'}{r^3} - \frac{2 H_0 p_3^2 \alpha \csc^2\theta f'}{r^3} - \frac{4 H_3 p_0^2 \alpha f'}{h r} \nonumber\\
&\quad + \frac{2 f H_2 p_2^2 \alpha h'}{h r^3} + \frac{2 f H_2 p_3^2 \alpha \csc^2\theta h'}{h r^3} + \frac{4 f^2 H_3 p_1^2 \alpha h'}{h r} \nonumber\\
&\quad + \frac{H_3 p_2^2 \alpha f' h'}{h r^2} + \frac{H_3 p_3^2 \alpha \csc^2\theta f' h'}{h r^2} - \frac{f H_3 p_2^2 \alpha (h')^2}{h^2 r^2} - \frac{f H_3 p_3^2 \alpha \csc^2\theta (h')^2}{h^2 r^2} \nonumber\\
&\quad + \frac{2 \mathcal{A}^\phi f p_0 p_1 q K_{XX} \varphi'}{h} - \mathcal{A}^\phi f^2 p_1^2 K_{XX} (\varphi')^2 
+ \frac{2 f H_3 p_2^2 \alpha h''}{h r^2} + \frac{2 f H_3 p_3^2 \alpha \csc^2\theta h''}{h r^2}.
\end{align}
\normalsize
Out of these a priori eleven equations, only two are independent.  In particular, comparing with Appendix B of Ref. \cite{Blazquez-Salcedo:2016enn} one can match ${}^{g}\mathcal{E}^{\text{lin}}_{12\,|_{\epsilon\rightarrow 0}}/{}^{g}\mathcal{E}^{\text{lin}}_{13\,|_{\epsilon\rightarrow 0}}$ to Eq. (B3), ${}^{g}\mathcal{E}^{\text{lin}}_{23\,|_{\epsilon\rightarrow 0}}$ to Eq. (B5), ${}^{g}\mathcal{E}^{\text{lin}}_{11\,|_{\epsilon\rightarrow 0}}$ to Eq. (B4), ${}^{g}\mathcal{E}^{\text{lin}}_{01\,|_{\epsilon\rightarrow 0}}$ to Eq. (B1), ${}^{g}\mathcal{E}^{\text{lin}}_{02\,|_{\epsilon\rightarrow 0}}/{}^{g}\mathcal{E}^{\text{lin}}_{03\,|_{\epsilon\rightarrow 0}}$ to Eq. (B2), ${}^{g}\mathcal{E}^{\text{lin}}_{00\,|_{\epsilon\rightarrow 0}}$ to Eq. (B6) and ${}^{\phi}\mathcal{E}^{\text{lin}}_{\,|_{\epsilon\rightarrow 0}}$ to Eq. (B7). We can use the redundant equations to eliminate from ${}^{\phi}\mathcal{E}^{\text{lin}}_{\,|_{\epsilon\rightarrow 0}}$ and ${}^{g}\mathcal{E}^{\text{lin}}_{00\,|_{\epsilon\rightarrow 0}}$ all other variables than $\mathcal{A}^\phi$ and $H_3$. We would like to point out the appearance of genuinely new terms owing to the presence of $q$.

\section{Components of the demixed system of characteristic equations}\label{app:DemixedEqsLO}
The four components of the matrix $Q^{IJ}(p)$ in Eq. \eqref{eq:charSys} read

\small
\begin{equation}
\begin{aligned}
Q^{\phi \phi} &= \frac{1}{h^3 r^4} \Bigg[
-32 f^4 h h' p_1^2 \varphi' r \alpha^2 \Big(G_{2XX} h (\varphi')^3 r^2 + 4 h' \alpha \Big) \\
&\quad + G_{2X} h r \Big(-p_0^2 r^2 + h (p_\Omega^2 + f p_1^2 r^2) \Big) (M_p^2 r - 8 f \varphi' \alpha) (h M_p^2 - 4 f h' \varphi' \alpha) \\
&\quad + h M_p^2 p_0^2 r \Big(-G_{2XX} M_p^2 q^2 r^3 + 64 f' h \alpha^2 \Big) \\
&\quad + 4 f^3 \alpha \Big[ G_{2XX} h p_1 (\varphi')^3 r^3 (2 h^2 M_p^2 p_1 + h h' M_p^2 p_1 r + 16 h' p_0 q \alpha) \\
&\quad + 8 h' \alpha \Big( h^2 M_p^2 p_1^2 r + 4 h (-h'' \varphi' p_\Omega^2 + h' \varphi'' p_\Omega^2 + h' p_1^2 \varphi' r - 2 p_0^2 \varphi'' r) \alpha \\
&\quad + 2 h' (h' \varphi' p_\Omega^2 - 4 p_0 p_1 q r) \alpha \Big) \Big] \\
&\quad + 2 f \Big[ G_{2XX} M_p^2 p_0 \varphi' q r^3 (h^2 M_p^2 p_1 r + 4 h p_0 q \alpha + 2 h' p_0 q r \alpha) \\
&\quad + 4 h \alpha^2 \Big(-4 h' M_p^2 p_0^2 r + 2 f' h M_p^2 (h' p_\Omega^2 - 4 p_0^2 r) + f' h' r (h' M_p^2 p_\Omega^2 - 16 p_0^2 \varphi' \alpha) \Big) \Big] \\
&\quad - f^2 \Big[ G_{2XX} (\varphi')^2 r^3 (h^3 M_p^4 p_1^2 r + 16 h^2 M_p^2 p_0 p_1 q \alpha + 8 h h' M_p^2 p_0 p_1 q r \alpha + 32 h' p_0^2 q^2 \alpha^2) \\
&\quad + 8 h' \alpha^2 \Big( 4 h^2 M_p^2 p_1^2 r + 2 h h' p_\Omega^2 (M_p^2 + 4 f' \varphi' \alpha) - 2 h r (M_p^2 (2 p_0^2 + h'' p_\Omega^2) + 8 p_0^2 (f' \varphi' + 2 \varphi'') \alpha) \\
&\quad + h' r (h' M_p^2 p_\Omega^2 - 32 p_0 p_1 q \alpha) \Big) \Big] \Bigg]\; ,
\end{aligned}
\end{equation}

\begin{equation}
\begin{aligned}
Q^{\phi g} &= \frac{1}{h^4 r^3} 
\Bigg[
\alpha \Big(
-4 h^3 M_p^4 p_0^2 r 
+ 64 f^4 h^2 (h')^2 p_1^2 (\varphi')^2 r \alpha^2 \\
&\quad - 16 f^3 h h' \varphi' \alpha \Big( 2 h' (h' \varphi' p_\Omega^2 + 4 p_0 p_1 q r) \alpha 
+ h (h' M_p^2 p_1^2 r^2 - 4 h'' \varphi' p_\Omega^2 \alpha + 8 p_0^2 \varphi'' r \alpha - 4 h' (\varphi'' p_\Omega^2 + p_1^2 \varphi' r) \alpha) \Big) \\
&\quad - 4 f^2 \Big( h^4 M_p^4 p_1^2 r - 16 (h')^3 p_\Omega^2 q^2 \alpha^2 + 2 h^2 h' r \alpha \big( M_p^2 (h'' \varphi' p_\Omega^2 + 4 p_0 p_1 q - 2 p_0^2 (\varphi' + 2 \varphi'' r)) + 16 p_0^2 \varphi' \varphi'' \alpha \big) \\
&\quad + h^3 M_p^2 \big( - h' M_p^2 p_1^2 r^2 + 4 h'' \varphi' p_\Omega^2 \alpha + 4 h' \varphi'' p_\Omega^2 \alpha + 8 h' p_1^2 \varphi' r \alpha + 8 p_0^2 \varphi'' r \alpha \big) \\
&\quad- h (h')^2 r \alpha \big( h' M_p^2 \varphi' p_\Omega^2 + 8 p_0 p_1 q (M_p^2 r - 4 \varphi' \alpha) \big) \Big) \\
&\quad + f' h^2 \Big( -2 h^2 M_p^4 p_\Omega^2 + h M_p^2 r (h' M_p^2 p_\Omega^2 - 16 (-1 + f) p_0^2 \varphi' \alpha) \\
&\quad+ 4 f h' \varphi' \alpha \big( h' p_\Omega^2 (- M_p^2 r + 8 f \varphi' \alpha) + 4 p_0^2 r (M_p^2 r - 4 (1+f) \varphi' \alpha) \big) \Big) \\
&\quad + f h^2 M_p^2 \Big( 4 h^2 M_p^2 p_1^2 r + 2 h \big( h' M_p^2 p_\Omega^2 + M_p^2 (2 p_0^2 + h'' p_\Omega^2) r + 16 p_0^2 \varphi'' r \alpha \big) \\
&\quad+ h' r \big( - h' M_p^2 p_\Omega^2 + 4 p_0 (- M_p^2 p_0 r + 4 p_0 \varphi' \alpha + 8 p_1 q \alpha) \big) \Big)
\Big)
\Bigg]\; , \\
\end{aligned}
\end{equation}

\begin{equation}
\begin{aligned}
Q^{g \phi} &= \frac{\alpha}{r^3 (h M_p^2 - 4 f h' \varphi' \alpha)} \Big[
4 h M_p^2 p_0^2 r 
- 16 f^3 h h' p_1^2 \varphi' r \alpha \\
&\quad + 4 f^2 \Big( h^2 M_p^2 p_1^2 r 
+ 4 h \big( h'' \varphi' p_\Omega^2 - h' \varphi'' p_\Omega^2 + h' p_1^2 \varphi' r + 2 p_0^2 \varphi'' r \big) \alpha 
- 2 h' \big( h' \varphi' p_\Omega^2 - 4 p_0 p_1 q r \big) \alpha \Big) \\
&\quad + f' h \Big( 2 h M_p^2 p_\Omega^2 + 16 (-1+f) p_0^2 \varphi' r \alpha - h' p_\Omega^2 (M_p^2 r + 8 f \varphi' \alpha) \Big) \\
&\quad - f \Big( 4 h^2 M_p^2 p_1^2 r + h' r \big( - h' M_p^2 p_\Omega^2 + 32 p_0 p_1 q \alpha \big) 
+ h \big( -2 h' M_p^2 p_\Omega^2 + 2 M_p^2 (2 p_0^2 + h'' p_\Omega^2) r + 32 p_0^2 \varphi'' r \alpha \big) \Big)
\Big]\; ,
\end{aligned}
\end{equation}

\begin{equation}
\begin{aligned}
Q^{gg} &= \frac{1}{2\, h\, r^2 \big( h M_p^2 - 4 f h' \varphi' \alpha \big)} \Big[
-16 f (h')^2 p_\Omega^2 q^2 \alpha^2 
+ 8 f h h' p_0 p_1 q r \alpha (- M_p^2 r + 8 f \varphi' \alpha) \\
&\quad + h^3 M_p^2 \Big( - M_p^2 (p_\Omega^2 + f p_1^2 r^2) 
+ 4 f' \varphi' p_\Omega^2 \alpha 
+ 8 f \big( \varphi'' p_\Omega^2 + f p_1^2 \varphi' r \big) \alpha \Big) \\
&\quad + h^2 \Big( M_p^4 p_0^2 r^2 
+ 4 M_p^2 \big( f h' \varphi' p_\Omega^2 - f' p_0^2 \varphi' r^2 + f^2 h' p_1^2 \varphi' r^2 - 2 f p_0^2 r (\varphi' + \varphi'' r) \big) \alpha \\
&\qquad - 16 f \varphi' \big( f' \varphi' (h' p_\Omega^2 - 2 p_0^2 r) 
+ 2 f (h' \varphi'' p_\Omega^2 + f h' p_1^2 \varphi' r - 2 p_0^2 \varphi'' r) \big) \alpha^2 \Big)
\Big]\; .
\end{aligned}
\end{equation}
\normalsize

\end{document}